\documentclass[aps,prd,onecolumn,preprintnumbers,notitlepage,superscriptaddress,
nofootinbib,amsfont,amssymb,10pt,singlespacing] {revtex4-1}
\usepackage{amsmath,bm}
\usepackage{times}
\usepackage{braket}
\usepackage{color,graphicx}
\usepackage{slashed}
\usepackage{hyperref}
\usepackage{graphics}
\usepackage{subfigure}
\usepackage{multirow,makecell}
\usepackage{textcomp}
\usepackage{mathtools}
\usepackage{float}

\newcommand{\beq}{\begin{eqnarray}}
\newcommand{\eeq}{\end{eqnarray}}

\def\lsim{ {\ \lower-1.2pt\vbox{\hbox{\rlap{$<$}\lower6pt\vbox{\hbox{$\sim$}
		}}}\ } }
\def\gsim{ {\ \lower-1.2pt\vbox{\hbox{\rlap{$>$}\lower6pt\vbox{\hbox{$\sim$}
		}}}\ } }
\definecolor{Red}{rgb}{1.,0.,0.}

\definecolor{Blue}{rgb}{0.,0.,1.}

\definecolor{nicered}{rgb}{0.7,0.1,0.1}
\definecolor{nicegreen}{rgb}{0.1,0.5,0.1}

\hypersetup{colorlinks,citecolor=nicegreen,linkcolor=nicered}

\begin{document}
	
	\title{Study of $B^{0}_{s} \rightarrow \ TT(a_{2}(1320),K^{*}_{2}(1430),f_{2}(1270),f^{'}_{2}(1525))$ in the perturbative QCD approach}
	
	\author{Jing~Dai}
	\email[Electronic address:]{2446041605@qq.com}
	\affiliation{School of Physical Science and Technology,
		Southwest University, Chongqing 400715, China}
	
	\author{Xian-Qiao~Yu}
	\email[Electronic address:]{yuxq@swu.edu.cn}
	\affiliation{School of Physical Science and Technology,
		Southwest University, Chongqing 400715, China}
	
	\date{\today}
	
	\begin{abstract}
		
		In the present study, the calculations of two-body decays  $B^{0}_{s} \rightarrow \ TT $ [ T denotes tensor mesons, $a_{2}(1320)$, $K^{*}_{2}(1430)$, $f_{2}(1270) $, $f^{'}_{2}(1525)$]  in the perturbative QCD approach are presented. The ensuing predictions encompass branching ratios, polarization fractions, and direct  $CP$ violations, all elucidated in comprehensive detail. It is discerned that (1) for pure annihilation decay, the longitudinal polarization is around  $90.0\%$, whereas the transverse polarizations manifest comparatively diminutive magnitudes. (2) The direct  $CP$ asymmetry is directly proportional to the interference between the tree and penguin contributions. For most of the decays investigated within this discourse, the direct  $CP$ asymmetry remains modest in magnitude. (3) There are precisely six distinct categories of Feynman diagrams for  $B^{0}_{s} \rightarrow \ TT $, because the tensor mesons cannot be produced through the  $ (V \pm A) $ currents or $ (S \pm P) $ density, thereby prohibiting factorizable emission diagrams. The nonfactorizable and annihilation contributions are ascertained to be pivotal in these decay modes. The calculated branching ratios of our calculation for  $B^{0}_{s} \rightarrow \ TT $ are at the order of $10^{-6}$ and $10^{-7}$, which can be tested in the LHCb and Belle II experiments.  (4) Mixing exists for the  $f_{2}(1270)$ and $f^{'}_{2}(1525)$, just as the  $\eta$ and $\eta^{'}$ mixing, the branching ratios about the mixing angle $\theta$ are given in this work. However, it is different from $f_{1}(1285)-f_{1}(1420)$, the mixing angle is notably small, thereby resulting in only marginal alterations in the decay branching ratios.

\end{abstract}

\maketitle

%
%

	\section{Introduction}\label{sec:intro}
The model elucidating the two-body decay of $ B $ mesons has been the subject of extensive scrutiny over the past two decades, both from a theoretical ~\cite{Liu:2017cwl,Kim:2013cpa,Zou:2015iwa,Ali:2007ff}and experimental~\cite{Munoz:1998sn,BaBar:2008ozy,BaBar:2008lan} perspective. The emergence of charmless hadronic $ B $ decays featuring a light tensor meson in the final state has invigorated interest in tensor mesons. Based on the principles of flavor SU(3) symmetry, we investigate nine mesons~\cite{PDG,Wang:2010ni}, comprising isovector mesons  $a_{2}(1320)$, isodoulet states $K^{*}_{2}(1430)$, and two isosinglet mesons $f_{2}(1270)$, $f^{'}_{2}(1525)$, which forms the first ${1}^{3}P_{2}$ nonet~\cite{ParticleDataGroup:2010dbb}. According to the latest experimental data of 2022 PDG~\cite{PDG}, the $B^{0}_{s} \rightarrow \ K^{*}_{2}(1430)\overline{\emph{K}}^{*}_{2}(1430)$ decay has already appeared, signifying the emergence of numerous decay modes involving the final state of two tensor mesons may appear. In Ref.~\cite{Kim:2013cpa}, some researchers have calculated the decay channel about $B^{0}_{s} \rightarrow \ VT $, at the same time, they also adopted some other methods to compare. Further, these calculations indicated that predictions based on the pQCD that can accommodate  experimental data well. By comparing their predictions with the experimental data, we find  that $  {\cal BR} (B^{0}_{s} \rightarrow \ \Phi(K^{*}_{2}(1430),\overline{\emph{K}}^{*}_{2}(1430) )) $ is similar to $ {\cal BR}( B^{0}_{s} \rightarrow \ \Phi(K^{*}(1430),\overline{\emph{K}}^{*}(1430) ))$, but the magnitude is smaller. For $B^{0}_{s} \rightarrow \ VV,TV $, they are similar but a little different.

In the exploration of two-body decay of $ B $ mesons, the perturbative QCD factorization approach assumes paramount significance. This  method based on factorization ~\cite{Yan:2018si,Li:2021qiw,Rui:2021kbn,Liang:2022mrz,Yang:2022jog,Liu:2021nun,Liang:2019eur} to calculate the decay process of $B^{0}_{s} \rightarrow \ M_{1}M_{2} $, in which $ M_{i} $ are composed of light noncharmed mesons. In the perturbative QCD framework, the factorization scale  about $ 1/b$ is employed to demarcate the boundary between the perturbative and nonperturbative regimes. The nonleptonic decay of the $B$ meson is postulated to be primarily governed by the exchange of hard gluons, permitting the isolation and direct computation of the hard portion of the decay process through perturbative methodologies. Simultaneously, the nonperturbative component is absorbed into the universal hadron wave function. On this foundation, the two-body decay amplitude of  $B^{0}_{s}$ meson is generically expressed as
\begin{equation}
{\cal A}={\cal H}\otimes \phi_{\emph{B}^{0}_{s}} \otimes \phi_{h_{1}} \otimes \phi_{h_{2}}.
\label{eq:exp1}
\end{equation}

Here the hard decay kernel ${\cal H}$ represents the contributions emanating from Feynman diagrams, amenable to computation through perturbative theory. The nonperturbative inputs  $\phi_{\emph{B}^{0}_{s}}$, $\phi_{h_{1}}$ and $\phi_{h_{2}}$ denote  the wave functions of $B^{0}_{s}$ meson, tensor mesons, respectively. These wave functions may be deduced via the extraction of pertinent empirical data or calculated through various nonperturbative methodologies.

Theoretical investigations into the calculations pertaining to tensor mesons have garnered the attention of several researchers. In comparison to vector mesons, tensor mesons are more special and complex. For $B \rightarrow \ PT $, $ VT $~\cite{Zou:2012td,Liu:2017cwl}, some scholars have investigated the tensor meson in the final state. However, for  $B^{0}_{s} \rightarrow \ TT $, the case where the final states are all tensor mesons has not yet been studied in the literature, and this article is the first  in this perspective. For the decays of $B^{0}_{s} \rightarrow \ TT $, the amplitude can be defined as three invariant helicity components: $A_{0}$, for which the  polarizations of the tensor meson are longitudinal with the respect to their momenta, and $A_{\Arrowvert}$, $A_{\perp}$ are for transversely polarized tensor meson~\cite{Huang:2005if,Li:2006cva}.

The branching ratios of  $B^{0}_{s}\rightarrow\ f_{2}(f^{'}_{2})f_{2}(f^{'}_{2})$ are contingent upon  the mixing angle $\theta$ of $f_{2}(1270)$ and $f^{'}_{2}(1525)$, analogous to $\eta$ and $\eta^{'}$ mixing in the pseudoscalar sector~\cite{Feldmann:1998vh,Xiao:2006mg,Ali:2007ff,Escribano:2005qq}. From the experiment that $\pi\pi$ is the dominant decay mode of $f_{2}(1270)$, and $f^{'}_{2}(1525)$ decays dominantly into $KK$, we can know that the physical $f_{2}(1270)$-$f^{'}_{2}(1525)$ mixing angle is smaller than the decoupling value: $\theta_{T,ph}-\theta_{dec.}$= $29.5^{\circ}-35.26^{\circ} =-5.8^{\circ} $~\cite{Cheng:2011fk,ParticleDataGroup:2010dbb}, which indicates  $f_{2}(1270)$ is nearly $f^{u}_{2}+f^{d}_{2}$, while $f^{'}_{2}(1525)$ is mainly $f^{s}_{2}$~\cite{Li:2000zb}.  The corresponding helicity amplitudes are characterized as follows~\cite{Jiang:2020eml,Yang:2007zt}

\begin{equation}
\begin{split}
f_{2}(1270) = \frac{1}{\sqrt{2}}(f^{u}_{2}+f^{d}_{2})\cos\theta_{f_{2}}-f^{s}_{2}\sin\theta_{f_{2}} ,\\
f^{'}_{2}(1525) = \frac{1}{\sqrt{2}}(f^{u}_{2}+f^{d}_{2})\cos\theta_{f_{2}}+f^{s}_{2}\sin\theta_{f_{2}} ,
\end{split}
\end{equation}
where $ f^{d}_{2} $ = $ d \overline d $, $ f^{u}_{2} $ = $ u \overline u $, $ f^{s}_{2} $ = $ s\overline s $ . Moreover, it is also found that the mixing angle  $ \theta_ {f_{2}} $ = $ 7.8 ^{\circ}$~\cite{Li:2000zb} and $ (9 \pm 1)^{\circ} $~\cite{Li:2002mi}.

This paper is structured as follows: In Sec. \ref{sec:pert}, we expound upon the theoretical underpinnings of the perturbative QCD (pQCD) framework and elaborate upon the wave functions integral to our calculation for the  $B^{0}_{s} \rightarrow \ TT $  decays. Section \ref{sec:amp} assembles the helicity amplitudes. Subsequently, in Sec. \ref{sec:numer}, we present the numerical results and engage in discussions. The key content of Sec. \ref{sec:summary}  comprises a summarization of the principal contributions of this study. Finally, the explicit formulations of all the helicity amplitudes are provided in the Appendix for reference.

\section{Theoretical Framework}\label{sec:pert}
\subsection{Hamiltonian and kinematics}
The pertinent weak effective Hamiltonian governing the decays  $B^{0}_{s} \rightarrow \ TT $  is defined by the following expression~\cite{Buras:1998raa,Buchalla:1995vs}
\begin{equation}
{\cal H}_{eff}=\frac{G_{F}}{\sqrt2}\big\{V^*_{ub}V_{us}[C_{1}O_{1}+C_{2}O_{2}]-V^*_{tb}V_{ts}[\sum^{10}_{i=3}C_{i}O_{i}]\big\},
\end{equation}
where $V^*_{ub}$$V_{us}$ and $V^{*}_{tb}$$V_{ts}$ are Cabibbo-Kobayashi-Maskawa factors, the Fermi coupling constant $G_{F}$=$1.66378\times10^{-5}{\rm GeV^{-2}}$, and $C_{i}$ is the Wilson coefficient corresponding to the quark operator, $O_{i}$ represents  the local four-quark operators, which can be expressed  as

\begin{equation}
\begin{split}\\
O_{1}&=\overline{b}_{\alpha}\gamma_{\mu}(1-\gamma_{5})u_{\beta}\overline{u}_{\beta}\gamma^{\mu}(1-\gamma_{5})X_{\alpha},\\
O_{2}&=\overline{b}_{\alpha}\gamma_{\mu}(1-\gamma_{5})u_{\alpha}\overline{u}_{\beta}\gamma^{\mu}(1-\gamma_{5})X_{\beta},\\
O_{3}&=\overline{b}_{\alpha}\gamma_{\mu}(1-\gamma_{5})X_{\alpha}\sum_{X^{'}}\overline{X^{'}}_{\beta}\gamma^{\mu}(1-\gamma_{5})X^{'}_{\beta},\\
O_{4}&=\overline{b}_{\alpha}\gamma_{\mu}(1-\gamma_{5})X_{\beta}\sum_{X^{'}}\overline{X^{'}}_{\beta}\gamma^{\mu}(1-\gamma_{5})X^{'}_{\alpha},\\
O_{5}&=\overline{b}_{\alpha}\gamma_{\mu}(1-\gamma_{5})X_{\alpha}\sum_{X^{'}}\overline{X^{'}}_{\beta}\gamma^{\mu}(1+\gamma_{5})X^{'}_{\beta},\\
O_{6}&=\overline{b}_{\alpha}\gamma_{\mu}(1-\gamma_{5})X_{\beta}\sum_{X^{'}}\overline{X^{'}}_{\beta}\gamma^{\mu}(1+\gamma_{5})X^{'}_{\alpha},\\
O_{7}&=\frac{3}{2}\overline{b}_{\alpha}\gamma_{\mu}(1-\gamma_{5})X_{\alpha}\sum_{X^{'}}e_{X^{'}}\overline{X^{'}}_{\beta}\gamma^{\mu}(1+\gamma_{5})X^{'}_{\beta},\\
O_{8}&=\frac{3}{2}\overline{b}_{\alpha}\gamma_{\mu}(1-\gamma_{5})X_{\beta}\sum_{X^{'}}e_{X^{'}}\overline{X^{'}}_{\beta}\gamma^{\mu}(1+\gamma_{5})X^{'}_{\alpha},\\
O_{9}&=\frac{3}{2}\overline{b}_{\alpha}\gamma_{\mu}(1-\gamma_{5})X_{\alpha}\sum_{X^{'}}e_{X^{'}}\overline{X^{'}}_{\beta}\gamma^{\mu}(1-\gamma_{5})X^{'}_{\beta},\\
O_{10}&=\frac{3}{2}\overline{b}_{\alpha}\gamma_{\mu}(1-\gamma_{5})X_{\beta}\sum_{X^{'}}e_{X^{'}}\overline{X^{'}}_{\beta}\gamma^{\mu}(1-\gamma_{5})X^{'}_{\alpha},
\end{split}
\end{equation}
where $\alpha$ and $\beta$ are color indices, ${X^{'}}=u,d,s,c$, or $b$ quarks, and they are the active quarks at the scale $ m_{b} $. $O_{1}$ and $O_{2}$ are current-current operators, $O_{i}(i=3,...,10)$ are penguin operators, in which $O_{i}(i=7,...,10)$ are the electroweak penguin operators. The operators $O_{7\gamma}$ and $O_{8g}$ are not listed, because their contribution is neglected.

Within the framework of the pQCD approach, the decay amplitude can be meticulously into three constituent components: the hard scattering kernel, the wave functions  characterizing the  mesons, and the convolution of the Wilson coefficients. For $B^{0}_{s}\rightarrow \ M_{1}M_{2}$ decay, the decay amplitude is presented as follows~\cite{Li:2006cva,Liu:2016rqu,Liu:2012jb,Li:2005hg}
\begin{equation}
{\cal A}\sim\int dx_{1}dx_{2}dx_{3}b_{1}db_{1}b_{2}db_{2}b_{3}db_{3}\cdot Tr[C(t)\Phi_{B}(x_{1},b_{1})\Phi_{M_{2}}(x_{2},b_{2})\Phi_{M_{3}}(x_{3},b_{3}) {\cal H} (x_{i},b_{i},t)
	S_{t} (x_{i}) e^{-S(t)}],
\end{equation}
where $ x_{i} $ are the proportions of the momenta for the spectator quark inside the mesons $B^{0}_{s}$, $T_{2}$ and $T_{3}$, respectively, with the values ranging from 0 to 1, $ b_{i} $ are the conjugate space coordinates of the transverse momenta $ k_{i} $ for the light quarks. $Tr$ denotes the trace over all Dirac structure and color indices. $C(t)$ is the short distance Wilson coefficients at the hard scale $ t $. $ t $ denotes the largest energy scale of the hard part $\cal H $. The threshold resummation $S_{t} (x_{i})$  stems from the large double logarithms~\cite{Li:2001ay}, which can remove the end point singularities. The last term $ e^{-S(t)} $ is the Sudakov factor, which can suppress soft dynamics~\cite{Li:1997un}.

In the context of the light cone coordinate system, the associated physical quantities are represented as follows. Assuming that the initial state of the meson $B^{0}_{s}$ is stationary, the tensor mesons $T_{2}$ and the $T_{3}$ move in the direction of the lightlike vector $v=(0,1,0_{\top})$ and $n=(1,0,0_{\top})$, respectively. Here we use ${p}_{1}$, ${p_{2}}$ and ${p}_{3}$ to represent the momenta of the mesons $B^0_{s}$, $T_{2}$ and $T_{3}$
\begin{equation}
\begin{split}
&\emph{p}_{1}=\frac{M_{B^{0}_{s}}}{\sqrt{2}}(1,1,0_{\top}),  \\
&\emph{p}_{2}=\frac{M_{B^{0}_{s}}}{\sqrt{2}}(1- r_{2}^2,r_{2}^2 ,0_{\top}),  \\
&\emph{p}_{3}=\frac{M_{B^{0}_{s}}}{\sqrt{2}}(r_{3}^2 ,1- r_{3}^2 ,0_{\top}).
\end{split}
\end{equation}

Moreover, the momenta of the respective light quarks associated with the mesons  $B^0_{s}$, $T_{2}$ and $T_{3}$ are

\begin{equation}
\begin{split}
&k_{1}=(0,\frac{M_{B^{0}_{s}}}{\sqrt{2}}x_{1},k_{1\top}),\\
&k_{2}=(\frac{M_{B^{0}_{s}}}{\sqrt{2}}(1- r_{2}^2 )x_{2},0,k_{2\top}),\\
&k_{3}=(0,\frac{M_{B^{0}_{s}}}{\sqrt{2}}(1- r_{3}^2)x_{3},k_{3\top}),
\end{split}
\end{equation}
where $ M_{B^{0}_{s}}$ represents the mass of the meson $B^{0}_{s}$, and $ r_{2}=\dfrac{M_{T_2}} {M_{B^{0}_{s}}}$, $ r_{3}=\dfrac{M_{T_3}} {M_{B^{0}_{s}}}$ ,
${M_{T}} $ is the mass of the tensor meson.

\subsection{Wave functions}
\subsubsection{B meson}
The wave function of the meson is expressed as a decomposition of Lorentz structures
 \begin{equation}
 \int\frac{d^{4}z}{(2\pi)^{4}}e^{ik\cdot z}<0\mid \overline b_{\alpha}(0)d_{\beta}(z)\mid B_{s}
( P_{1})>\\
=\frac{i}{\sqrt{2N_{c}}}(\not {p}_{1}+M_{B}){\gamma_{5}}[\phi_{B}({x_{1},b_{1}})+\frac{\not {n}}{\sqrt{2}} \overline{\phi}_{B}(x_{1},b_{1})],
 \end{equation}\\
where $ \phi_{B}({x_{1},b_{1}}) $and $ \overline{\phi}_{B}({x_{1},b_{1}}) $ are the twist distribution amplitudes, the contribution of $ \overline{\phi}_{B}({x_{1},b_{1}}) $ is relatively small, so we neglect it. Therefore, the meson $B^{0}_{s}$ is deemed to be a heavy-light model, with the wave function defined as~\cite{Shen:2014wga,Keum:2000ph,Beneke:2000wa,Lu:2000em}

\begin{equation}
\Phi_{B}=\frac{i}{\sqrt{2N_{c}}}(\not {p}_{1}+M_{B}){\gamma_{5}}{\phi_{B}({x_{1},b_{1}})},
\end{equation}
where $N_{c}={3}$ is the number of colors, because we calculate the relevant parameters in the $ b $ space, and the distribution amplitude $\phi_{B}$ can be expressed as~\cite{Kurimoto:2001zj,Li:2003yj}
\begin{equation}
\phi_{B}(x_{1},b_{1})=\emph{N}_{B}{{x_{1}}^2}(1-{x_{1}})^2\exp[-\frac{M^2_{B}{{x_{1}}^2}}{2\omega^2_{B}}-\frac{1}{2}(\omega_{B}{b_{1}})^2].
\end{equation}

This distribution amplitude adheres to the normalization condition
\begin{equation}
\int^{1}_{0}dx\phi_{B}(x,b=0)=\frac{f_{B}}{{2}\sqrt{2N_{c}}},
\end{equation}
where $N_{B}=91.784$ GeV is the normalization constant, $f_{B}$ is the decay constant. For  $B^{0}_{s}$ meson, we use the shape parameter $\omega_{B_{s}}=0.50\pm0.05$ GeV~\cite{Ali:2007ff}.\\

\subsubsection{Tensor meson}
For the spin-2 polarization tensor $\epsilon^{uv}(\lambda) $ with helicity $ \lambda $, satisfies $ \epsilon_{uv}p^{v}_{2} $ = 0~\cite{Cheng:2010hn,Cheng:2010yd}, which can be constructed based on the polarization vectors of vector mesons $\epsilon$, they can be written as
\begin{equation}
\begin{split}
\epsilon_{uv}(\pm2) &\equiv  \epsilon(\pm1)_{u}  \epsilon(\pm1)_{v} ,\\
\epsilon_{uv}(\pm1) &\equiv \frac{1}{\sqrt{2}} [\epsilon(\pm1)_{u} \epsilon(0)_{v}+ \epsilon(0)_{u}  \epsilon(\pm1)_{v} ],\\
\epsilon_{uv}(0) &\equiv \frac{1}{\sqrt{6}}[\epsilon(+1)_{u} \epsilon(-1)_{v}+\epsilon(-1)_{u}  \epsilon(+1)_{v}]+\sqrt\frac{2}{{3}}\epsilon(0)_{u}  \epsilon(0)_{v}.\\
\end{split}
\end{equation}

With the tensor meson  moving in the plus direction of the $ z $ axis, the polarizations $ \epsilon $ are  defined as
\begin{equation}
\begin{split}
\epsilon_{u}(0)=\frac{1}{m_{T}}(\mid P_{T}\mid,0,0, E_{T}),\\
\epsilon_{u}(\pm1)=\frac{1}{\sqrt{2}}(0,\pm1,i,0),
\end{split}
\end{equation}
where $E_{T}$ represents the energy of the tensor meson. In the subsequent calculations, the introduction of a new polarization vector $ \epsilon_{T} $  for the tensor meson under consideration is deemed necessary for the sake of convenience~\cite{Wang:2010ni}
\begin{equation}
\begin{split}
\epsilon_{Tu}(\lambda)=\frac{1}{m_{B}}\epsilon_{uv}(\lambda)P^{v}_{B},\\
\end{split}
\end{equation}
which satisfies
\begin{equation}
\begin{split}
\epsilon_{Tu}(\pm2)&=0,\\
\epsilon_{Tu}(\pm1)&=\frac{1}{\sqrt{2}m_{B}}\epsilon(0)\cdot P_{B}\epsilon_{u}(\pm1),\\
\epsilon_{Tu}(\pm0)&=\sqrt{\frac{2}{3}}\frac{1}{m_{B}}\epsilon(0)\cdot P_{B}\epsilon_{u}(0).\\
\end{split}
\end{equation}

The contraction is evaluated as $ \epsilon(0)\cdot P_{B}/m_{B} = \mid\overrightarrow P_{T}\mid/m_{T}$. It is obvious that the new vector $ \epsilon_{T} $ is similar to the ordinary polarization vector $ \epsilon $, regardless of the dimensionless constants $\sqrt{\frac{1}{2}}\frac{\mid \overrightarrow P_{T}\mid}{m_{T}} $ or $\sqrt{\frac{2}{3}}\frac{\mid \overrightarrow P_{T}\mid}{m_{T}} $.

The decay constants of the tensor mesons are defined as~\cite{Wang:2010ni}
\begin{equation}
\begin{split}
<T\mid j_{uv}(0)\mid0>=f_{T}m^{2}_{T}\epsilon^{*}_{uv},\\
<T\mid j_{uv\rho}\mid0>=-if^{T}_{T}m_{T} (\epsilon^{*}_{u\rho}P_{2v}-\epsilon^{*}_{v\rho}P_{2u}).
\end{split}
\end{equation}

Where the currents are expressed as
\begin{equation}
\begin{split}
 j_{uv}(0)&=\frac{1}{2}[\overline q_{1}(0)\gamma_{u}i\overleftrightarrow D_{v}q_{2}(0)+\overline q_{1}(0)\gamma_{v}i\overleftrightarrow D_{u}q_{2}(0)],\\
 j^{+}_{uv\rho}(0)&=\overline q_{2}(0)\sigma_{uv}i\overleftrightarrow D_{\rho}q_{1}(0),
\end{split}
\end{equation}
with$\overleftrightarrow D_{u}= \overrightarrow D_{u}-\overleftarrow  D_{u} $, $\overrightarrow D_{u}=\overrightarrow \partial_{u}+ig_{s}A^{a}_{u}\lambda^{a}/{2} $ and $\overleftarrow D_{u} =\overleftarrow \partial _{u}-ig_{s}A^{a}_{u}\lambda^{a}/{2} $, respectively. Here we adopted these decay constants from Ref.~\cite{Cheng:2010hn} that have been calculated in the QCD sum rules~\cite{Aliev:1981ju,Aliev:1982ab,Aliev:2009nn}, which can be seen from Table \ref{111}. We can find that the transverse decay constants are approximately equal to the longitudinal one for $ a_{2}(1320) $, $ f_{2}(1270) $, but it is different from $K^{*}_{2}(1430)$ and $ f^{'}_{2}(1525) $, their ratio relationship: $ f^{T}_{T} /f_{T} \sim (50\%-65\%)$.
\begin{table}[htbp]
	\centering
	\caption{Decay constants (in unit of MeV) of tensor mesons}
	\label{111}
	\begin{tabular*}{\columnwidth}{@{\extracolsep{\fill}}lllll@{}}
		\hline
		\hline
		\\
	  &$f_{a_{2}(1320)}=107\pm 6$
		&$ f_{K^{*}_{2}(1430)}=118\pm 5$  	&$f_{f_{2}(1270)}=102\pm 6$   &$ f_{f^{'}_{2}(1525)}=126\pm 4$ \\
	\hline
	\\
	     &$f^{T}_{a_{2}(1320)}=105\pm 21$
		&$ f^{T}_{K^{*}_{2}(1430)}=77\pm 14$  &$f^{T}_{f_{2}(1270)}=117\pm 25$
		&$ f^{T}_{f^{'}_{2}(1525)}=65\pm 12$ \\
		\hline
		\hline
	\end{tabular*}
\end{table}

 From earlier studies ~\cite{Cheng:2010hn}, we  obtain insights into the light cone distribution amplitudes of the tensor mesons. The light cone distribution amplitudes up to twist-3 for generic tensor  mesons are defined as follows
\begin{equation}
\begin{split}\label{equ:d}
<T(p_{2},\epsilon)\mid {q_{1\alpha}(0)\overline q_{2\beta}(z)\mid 0}>
=  \frac{1}{\sqrt{2N_{c}} }
\int^{1}_{0}\emph{d}\emph{x} e^{ixp_{2} \cdot {z}}[m_{T}\not {\epsilon}^{*}_{\bullet L}\Phi_{T}(x)\\
+\not {\epsilon}^{*}_{\bullet L}\not {p_{2}}\Phi^{t}_{T}(x)+m^{2}_{T}\frac{\epsilon_{\bullet}\cdot {v}}{p_{2}\cdot{v}}\Phi^{s}_{T}(x)]_{\alpha\beta},
\end{split}
\end{equation}
\begin{equation}
\begin{split}\label{equ:r}
<T(p_{2},\epsilon)\mid {q_{1\alpha}(0)\overline q_{2\beta}(z)\mid 0}>
=\frac{1}{\sqrt{2N_{c}}} \int^{1}_{0}\emph{d}\emph{x} e^{ixp_{2}\cdot{z}}[m_{T}\not {\epsilon}^{*}_{\bullet T}\Phi^{v}_{T}(x)\\
+\not {\epsilon}^{*}_{\bullet T}\not {p_{2}}\Phi^{T}_{T}(x)+m_{T}i\epsilon_{uv\rho\sigma}\gamma_{5}\gamma^{u}\epsilon^{*v}_{\bullet T}n^{\rho}v^{\sigma}\Phi^{a}_{T}(x)]_{\alpha\beta}.
\end{split}
\end{equation}

 The convention $\epsilon^{0123} $=1 has been adopted. Equation~(\ref{equ:d}) is for the longitudinal polarization$( \lambda= 0 )$, and Eq.~(\ref{equ:r}) is for the transverse polarizations$( \lambda =\pm{1})$, respectively. Here $n$ is the moving direction of the tensor meson and $v$ is the opposite direction. The new vector $ \epsilon_{\bullet} $ which plays the same role with the polarization vector $ \epsilon $, which is defined by
 \begin{equation}
 \begin{split}
 \epsilon_{\bullet u}= \frac{\epsilon_{uv}v^{v}}{p_{2}\cdot{v}} m_{T}.
 \end{split}
 \end{equation}

 With the momenta and polarizations, which can be reexpressed as
 \begin{equation}
 \begin{split}
 \epsilon_{\bullet u} = \frac{2m_{T}}{m^{2}_{B}} p^{v}_{B}\epsilon_{uv}.
 \end{split}
 \end{equation}

  In earlier studies~\cite{Wang:2010ni,Cheng:2010hn,Cheng:2010yd}, the amplitudes are expressed as
 \begin{equation}
 \begin{split}
 \Phi_{T}(x)&=\frac{f_{T}}{2\sqrt{2N_{c}}}\Phi_{\parallel}(x),\Phi^{t}_{T}(x)=\frac{f^{T}_{T}}{2\sqrt{2N_{c}}}h^{(t)}_{\parallel}(x),\\
 \Phi^{s}_{T}(x)&=\frac{f^{T}_{T}}{4\sqrt{2N_{c}}}\frac{d}{dx}h^{s}_{\parallel}(x),\Phi^{T}_{T}(x)=\frac{f^{T}_{T}}{2\sqrt{2N_{c}}}\Phi_{\perp}(x),\\
 \Phi^{v}_{T}(x)&=\frac{f_{T}}{2\sqrt{2N_{c}}}g^{v}_{\perp}(x),\Phi^{a}_{T}(x)=\frac{f_{T}}{8\sqrt{2N_{c}}}\frac{d}{dx}g^{a}_{\perp}(x).\\
 \end{split}
 \end{equation}

 The twist-2 distribution amplitudes can be expanded in terms of Gegenbauer polynomials $ C^{3/2}_{n}(2x-1) $, with the asymptotic form  given by
 \begin{equation}
 \begin{split}
 \Phi_{\parallel,\perp}(x)=30x(1-x)(2x-1).
 \end{split}
 \end{equation}

 Adhering to normalization conditions
 \begin{equation}
 \begin{split}
 \int^{1}_{0}\emph{d}\emph{x}(2x-1) \Phi_{\parallel,\perp}(x)=1.
 \end{split}
 \end{equation}

 The twist-3 distribution amplitudes also assume an asymptotic form, as delineated in~\cite{Cheng:2010hn}
 \begin{equation}
 \begin{split}
 h^{t}_{\parallel}(x) &=\frac{15}{2}  (2x-1)(1-6x+6x^{2}) ,\\
 h^{s}_{\parallel}(x) &= 15x(1-x)(2x-1) ,\\
 g^{a}_{\perp}(x) &= 20x(1-x)(2x-1) ,\\
 g^{v}_{\perp}(x) &= 5(2x-1)^{3}.
 \end{split}
 \end{equation}

\section{Decay amplitudes}\label{sec:amp}
\begin{figure}[htbp]
	\centering
	\begin{tabular}{l}
		\includegraphics[width=0.8\textwidth]{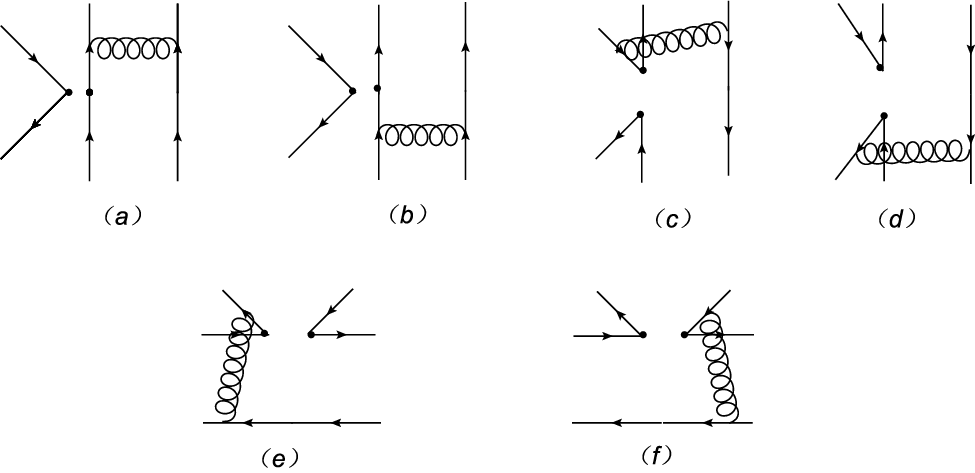}
	\end{tabular}
	\caption {The Feynman diagrams for the $B^{0}_{s} \rightarrow \ TT $ decays }
	\label{fig1}
\end{figure}

In this section, we provide the perturbative QCD formulas for all the Feynman diagrams, as illustrated  in Fig.~\ref{fig1}. The first row showcases the annihilation-type diagrams, with the first two  being factorizable and the last two being nonfactorizable. The second row consists of nonfactorizable emission diagrams. For the $B^{0}_{s} \rightarrow \ TT $ decays, both the longitudinal polarization and the transverse polarization contribute. The symbol $\emph{F}$ and $\emph{M}$ represent the factorizable and nonfactorizable contributions, respectively. The superscripts $\emph{LL}$ denotes the amplitude of the $ (V-A)(V-A) $ operators, and $\emph{LR}$ describe the amplitude of the $ (V-A)(V+A) $ operators. The symbol $\emph{SP}$ is Fierz transformation of $\emph{LR}$. Notably, the decay amplitudes for longitudinal and transverse polarizations exhibit the same form after simplification, as follows in Eq.~(\ref{equ:g})-(\ref{equ:h}).\\

The longitudinal polarization amplitudes of the factorizable annihilation diagrams are
\begin{equation}
\begin{split}\label{equ:g}
{\emph F}^{\emph{LL,L}}_{af}=&\frac{16}{3}\pi\emph{C}_{\emph{F}}f_{B}M^{4}_{B^0_{s}}\int^{1}_{0}\emph{d}\emph{x}_{2}\emph{d}\emph{x}_{3}\int^\frac{1}{\Lambda}_{0}\emph{b}_{2}\emph{d}\emph{b}_{2}\emph{b}_{3}\emph{d}\emph{b}_{3}\\
& \times \{[2{r}_{2}{r}_{3}{x}_{3}\phi^{s}_{T}({x}_{2})\phi^{t}_{T}({x}_{3})-2{r}_{2}{r}_{3}{x}_{3} ({x}_{3}-2)\phi^{s}_{T}({x}_{2})\phi^{s}_{T}({x}_{3})+({x}_{3}-1)\phi_{T}({x}_{2})\phi_{T}({x}_{3})]{E}_{af}({t}_{e}){h}_{af}(\alpha_{1},\beta,{b}_{2},{b}_{3})\\
&+[-2{r}_{2}{r}_{3}({x}_{2}-1)\phi^{s}_{T}({x}_{3})\phi^{t}_{T}({x}_{2})+2{r}_{2}{r}_{3}(-{x}_{2}-1)\phi^{s}_{T}({x}_{2})\phi^{s}_{T}({x}_{3})+{x}_{2}\phi_{T}({x}_{2})\phi_{T}({x}_{3})]{E}_{af}({t}_{f}){h}_{af}(\alpha_{2},\beta,{b}_{3},{b}_{2})\} \\
\end{split}
\end{equation}
\begin{equation}
\begin{split}
{\emph F}^{\emph{SP,L}}_{af}=&-\frac{32}{3}\pi\emph{C}_{\emph{F}}f_{B}M^{4}_{B^0_{s}}\int^{1}_{0}\emph{d}\emph{x}_{2}\emph{d}\emph{x}_{3}\int^\frac{1}{\Lambda}_{0}\emph{b}_{2}\emph{d}\emph{b}_{2}\emph{b}_{3}\emph{d}\emph{b}_{3}\\
& \times \{[2{r}_{2}\phi^{s}_{T}({x}_{2})\phi_{T}({x}_{3})+{r}_{3}({x}_{3}-1)\phi^{s}_{T}({x}_{3})\phi_{T}({x}_{2})+{r}_{3}({x}_{3}-1)\phi_{T}({x}_{2})\phi^{t}_{T}({x}_{3})]{E}_{af}({t}_{e}){h}_{af}(\alpha_{1},\beta,{b}_{2},{b}_{3})\\
& +[2{r}_{3}({x}_{2}-1)\phi^{s}_{T}({x}_{3})\phi_{T}({x}_{2})+{r}_{2}{x}_{2}\phi^{s}_{T}({x}_{3})\phi^{t}_{T}({x}_{2})-{r}_{2}{x}_{2}\phi^{s}_{T}({x}_{2})\phi_{T}({x}_{3})]{E}_{af}({t}_{f}){h}_{af}(\alpha_{2},\beta,{b}_{3},{b}_{2}) \}\\
\end{split}
\end{equation}

The longitudinal polarization amplitudes of the nonfactorizable annihilation diagrams are given below
\begin{equation}
\begin{split}
{\emph M}^{\emph{LL,L}}_{anf}=&\frac{32}{3}\sqrt\frac{2}{3}\pi\emph{C}_{\emph{F}}M^{4}_{B^0_{s}}\int^{1}_{0}\emph{d}\emph{x}_{1}\emph{d}\emph{x}_{2}\emph{d}\emph{x}_{3}\int^\frac{1}{\Lambda}_{0}\emph{b}_{1}\emph{d}\emph{b}_{1}\emph{b}_{2}\emph{d}\emph{b}_{2}\phi_{B_{s}}(\emph{x}_{1},b_{1})\\
& \times \{[{r}_{2}{r}_{3}((1-{x}_{2}+{x}_{3}) \phi^{t}_{T}({x}_{2})\phi^{t}_{T}({x}_{3})+({x}_{2}+{x}_{3}-1)\phi^{t}_{T}({x}_{2})\phi^{s}_{T}({x}_{3})\\
&+(1-{x}_{2}-{x}_{3})\phi_{T}({x}_{2})\phi_{T}({x}_{3})+({x}_{2}-{x}_{3}+3)\phi^{s}_{T}({x}_{2})\phi^{s}_{T}({x}_{3}))\\
&-{x}_{2}\phi_{T}({x}_{2})\phi_{T}({x}_{3})]{E}_{anf}({t}_{g}){h}_{anf}(\alpha,\beta_{1},{b}_{1},{b}_{2})\\
&+[-{r}_{2}{r}_{3}((1+{x}_{2}-{x}_{3})\phi^{s}_{T}({x}_{2})\phi^{s}_{T}({x}_{3})+({x}_{2}+{x}_{3}-1)\phi^{s}_{T}({x}_{2})\phi^{t}_{T}({x}_{3})\\
&+(1-{x}_{2}-{x}_{3})\phi^{t}_{T}({x}_{2})\phi^{s}_{T}({x}_{3})+({x}_{3}-{x}_{2}-1)\phi^{t}_{T}({x}_{2})\phi^{t}_{T}({x}_{3}))\\
&+({x}_{3}-1)\phi_{T}({x}_{2})\phi_{T}({x}_{3})]{E}_{anf}({t}_{h}){h}_{anf}(\alpha,\beta_{2},{b}_{1},{b}_{2}) \} \\
\end{split}
\end{equation}
\begin{equation}
\begin{split}
{\emph M}^{\emph{LR,L}}_{anf}=&\frac{32}{3}\sqrt\frac{2}{3}\pi\emph{C}_{\emph{F}}M^{4}_{B^0_{s}}\int^{1}_{0}\emph{d}\emph{x}_{1}\emph{d}\emph{x}_{2}\emph{d}\emph{x}_{3}\int^\frac{1}{\Lambda}_{0}\emph{b}_{1}\emph{d}\emph{b}_{1}\emph{b}_{2}\emph{d}\emph{b}_{2}\phi_{B_{s}}(\emph{x}_{1},b_{1})\\
& \times \{[{r}_{2}(2-{x}_{2}) \phi^{s}_{T}({x}_{2})\phi_{T}({x}_{3})+ {r}_{2}(2-{x}_{2})\phi^{t}_{T}({x}_{2})\phi_{T}({x}_{3})\\
&+{r}_{3}({x}_{3}+1)\phi_{T}({x}_{2})\phi^{s}_{T}({x}_{3})-{r}_{3}({x}_{3}+1) \phi^{t}_{T}({x}_{3})\phi_{T}({x}_{2})]{E}_{anf}({t}_{g}){h}_{anf}(\alpha,\beta_{1},{b}_{1},{b}_{2})\\
& +[{r}_{2}x_{2}(\phi^{s}_{T}({x}_{2})\phi_{T}({x}_{3})+\phi^{t}_{T}({x}_{2})\phi_{T}({x}_{3}))+{r}_{3}(1-x_{3})(\phi_{T}({x}_{2})\phi^{s}_{T}({x}_{3})\\
&-\phi_{T}({x}_{2})\phi^{t}_{T}({x}_{3}))]{E}_{anf}({t}_{h}){h}_{anf}(\alpha,\beta_{2},{b}_{1},{b}_{2})\} \\
\end{split}
\end{equation}
\begin{equation}
\begin{split}
{\emph M}^{\emph{SP,L}}_{anf}=&-\frac{32}{3}\sqrt\frac{2}{3}\pi\emph{C}_{\emph{F}}f_{B}M^{4}_{B^0_{s}}\int^{1}_{0}\emph{d}\emph{x}_{1}\emph{d}\emph{x}_{2}\emph{d}\emph{x}_{3}\int^\frac{1}{\Lambda}_{0}\emph{b}_{1}\emph{d}\emph{b}_{1}\emph{b}_{2}\emph{d}\emph{b}_{2}\phi_{B_{s}}(\emph{x}_{1},b_{1})\\
& \times \{[{r}_{2}{r}_{3}((1-{x}_{2}+{x}_{3})\phi^{t}_{T}({x}_{2})\phi^{t}_{T}({x}_{3})-({x}_{2}+{x}_{3}-1)\phi^{t}_{T}({x}_{2})\phi^{s}_{T}({x}_{3})\\
&+({x}_{2}+{x}_{3}-1)\phi^{s}_{T}({x}_{2})\phi^{t}_{T}({x}_{3})+({x}_{2}-{x}_{3}+3)\phi^{s}_{T}({x}_{2})\phi^{s}_{T}({x}_{3})\\
&+({x}_{3}-1)\phi_{T}({x}_{2})\phi_{T}({x}_{3})]{E}_{anf}({t}_{g}){h}_{anf}(\alpha,\beta_{1},{b}_{1},{b}_{2})\\
&-[{r}_{2}{r}_{3}((1+{x}_{2}-{x}_{3})\phi^{s}_{T}({x}_{2})\phi^{s}_{T}({x}_{3})+(1-{x}_{2}-{x}_{3})\phi^{s}_{T}({x}_{2})\phi^{t}_{T}({x}_{3})+({x}_{2}+{x}_{3}-1)\phi^{t}_{T}({x}_{2})\phi^{s}_{T}({x}_{3})\\
&+({x}_{3}-{x}_{2}-1)\phi^{t}_{T}({x}_{2})\phi^{t}_{T}({x}_{3}))-{x}_{2}\phi_{T}({x}_{2})\phi_{T}({x}_{3})]{E}_{anf}({t}_{h}){h}_{anf}(\alpha,\beta_{2},{b}_{1},{b}_{2}) \} \\
\end{split}
\end{equation}

The longitudinal polarization amplitudes of the nonfactorizable emission diagrams are as follows
\begin{equation}
\begin{split}
{\emph M}^{\emph{LL,L}}_{enf}=-&\frac{32}{3}\sqrt\frac{2}{3}\pi\emph{C}_{\emph{F}}M^{4}_{B^0_{s}}\int^{1}_{0}\emph{d}\emph{x}_{1}\emph{d}\emph{x}_{2}\emph{d}\emph{x}_{3}\int^\frac{1}{\Lambda}_{0}\emph{b}_{1}\emph{d}\emph{b}_{1}\emph{b}_{2}\emph{d}\emph{b}_{2}\phi_{B_{s}}(\emph{x}_{1},b_{1})\phi_{2}({x}_{2})\\
& \times \{[({x}_{2}-1) \phi_{T}({x}_{3})+ {r}_{3}({x}_{3}-1)\phi^{s}_{T}({x}_{3})-{r}_{3}({x}_{3}-1)\phi^{t}_{T}({x}_{3})]{E}_{enf}({t}_{c}){h}_{enf}(\alpha,\beta_{1},{b}_{1},{b}_{2})\\
& +[(x_{2}+x_{3}-2)\phi_{T}({x}_{3})-{r}_{3}(x_{3}-1)\phi^{s}_{T}({x}_{3})-{r}_{3}(x_{3}-1)\phi^{t}_{T}({x}_{3})]{E}_{enf}({t}_{d}){h}_{enf}(\alpha,\beta_{2},{b}_{1},{b}_{2})\} \\
\end{split}
\end{equation}

\begin{equation}
\begin{split}
{\emph M}^{\emph{LR,L}}_{enf}=&\frac{32}{3}\sqrt\frac{2}{3}\pi\emph{C}_{\emph{F}}M^{4}_{B^0_{s}}{r}_{2}\int^{1}_{0}\emph{d}\emph{x}_{1}\emph{d}\emph{x}_{2}\emph{d}\emph{x}_{3}\int^\frac{1}{\Lambda}_{0}\emph{b}_{1}\emph{d}\emph{b}_{1}\emph{b}_{2}\emph{d}\emph{b}_{2}\phi_{B_{s}}(\emph{x}_{1},b_{1})\\
& \times \{[{r}_{3}({x}_{2}-{x}_{3}-1) (\phi^{s}_{T}({x}_{2})\phi^{s}_{T}({x}_{3})-\phi^{t}_{T}({x}_{2})\phi^{t}_{T}({x}_{3}))+ ({x}_{2}+{x}_{3}-1)(\phi^{s}_{T}({x}_{3})\phi^{t}_{T}({x}_{2})-\phi^{s}_{T}({x}_{2})\phi^{s}_{T}({x}_{3}))\\
&+({x}_{2}-1)(\phi_{T}({x}_{3})\phi^{s}_{T}({x}_{2})-\phi_{T}({x}_{3})\phi^{t}_{T}({x}_{2}))]{E}_{enf}({t}_{c}){h}_{enf}(\alpha,\beta_{1},{b}_{1},{b}_{2})\\
&+[{r}_{3}(x_{3}-{x}_{2})(\phi^{t}_{T}({x}_{2})\phi^{s}_{T}({x}_{3})+\phi^{s}_{T}({x}_{2})\phi^{t}_{T}({x}_{3}))+({x}_{2}+{x}_{3})(\phi^{s}_{T}({x}_{2})\phi^{s}_{T}({x}_{3})+\phi^{t}_{T}({x}_{2})\phi^{t}_{T}({x}_{3}))\\
&+{x}_{2}(\phi_{T}({x}_{3})\phi^{s}_{T}({x}_{2})-\phi_{T}({x}_{3})\phi^{t}_{T}({x}_{2}))]{E}_{enf}({t}_{d}){h}_{enf}(\alpha,\beta_{2},{b}_{1},{b}_{2}) \} \\
\end{split}
\end{equation}
\begin{equation}
\begin{split}
{\emph M}^{\emph{SP,L}}_{enf}=&-\frac{32}{3}\sqrt\frac{2}{3}\pi\emph{C}_{\emph{F}}M^{4}_{B^0_{s}}\int^{1}_{0}\emph{d}\emph{x}_{1}\emph{d}\emph{x}_{2}\emph{d}\emph{x}_{3}\int^\frac{1}{\Lambda}_{0}\emph{b}_{1}\emph{d}\emph{b}_{1}\emph{b}_{2}\emph{d}\emph{b}_{2}\phi_{B_{s}}(\emph{x}_{1},b_{1})\phi_{2}({x}_{2})\\
& \times \{[({x}_{2}-{x}_{3}-1) \phi_{T}({x}_{3})+ {r}_{3}{x}_{3}\phi^{s}_{T}({x}_{3})+{r}_{3}{x}_{3}\phi^{t}_{T}({x}_{3})]{E}_{enf}({t}_{c}){h}_{enf}(\alpha,\beta_{1},{b}_{1},{b}_{2})\\
&+[-{r}_{3}{x}_{3}\phi^{s}_{T}({x}_{3})+{r}_{3}{x}_{3}\phi^{t}_{T}({x}_{3})+{x}_{2}\phi_{T}({x}_{3})]{E}_{enf}({t}_{d}){h}_{enf}(\alpha,\beta_{2},{b}_{1},{b}_{2}) \}\\
\end{split}
\end{equation}

The transverse polarization amplitudes of the factorizable annihilation diagrams are
\begin{equation}
\begin{split}
{\emph F}^{\emph{LL(LR),N}}_{af}=&4\pi\emph{C}_{\emph{F}}f_{B}M^{4}_{B^0_{s}}{r}_{2}{r}_{3}\int^{1}_{0}\emph{d}\emph{x}_{2}\emph{d}\emph{x}_{3}\int^\frac{1}{\Lambda}_{0}\emph{b}_{2}\emph{d}\emph{b}_{2}\emph{b}_{3}\emph{d}\emph{b}_{3}\\
& \times \{[({x}_{3}-2)(\phi^{v}_{T}({x}_{2})\phi^{v}_{T}({x}_{3})+ \phi^{a}_{T}({x}_{2})\phi^{a}_{T}({x}_{3}))-{x}_{3}\phi^{v}_{T}({x}_{2})\phi^{a}_{T}({x}_{3})\\
&-{x}_{3}\phi^{a}_{T}({x}_{2})\phi^{v}_{T}({x}_{3})]{E}_{af}({t}_{e}){h}_{af}(\alpha_{1},\beta,{b}_{2},{b}_{3})\\
&+[({x}_{2}-1)\phi^{v}_{T}({x}_{3})\phi^{a}_{T}({x}_{2})+({x}_{2}-1)\phi^{a}_{T}({x}_{3})\phi^{v}_{T}({x}_{2})\\
&+({x}_{2}+1)\phi^{a}_{T}({x}_{3})\phi^{a}_{T}({x}_{2})+({x}_{2}+1)\phi^{v}_{T}({x}_{3})\phi^{v}_{T}({x}_{2})]{E}_{af}({t}_{f}){h}_{af}(\alpha_{2},\beta,{b}_{3},{b}_{2}) \}\\
\end{split}
\end{equation}

\begin{equation}
\begin{split}
{\emph F}^{\emph{SP,N}}_{af}=&8\pi\emph{C}_{\emph{F}}f_{B}M^{4}_{B^0_{s}}\int^{1}_{0}\emph{d}\emph{x}_{2}\emph{d}\emph{x}_{3}\int^\frac{1}{\Lambda}_{0}\emph{b}_{2}\emph{d}\emph{b}_{2}\emph{b}_{3}\emph{d}\emph{b}_{3}\\
& \times \{[{r}_{2}\phi^{a}_{T}({x}_{2})\phi^{T}_{T}({x}_{3})+{r}_{2}\phi^{v}_{T}({x}_{2})\phi^{T}_{T}({x}_{3})]{E}_{af}({t}_{e}){h}_{af}(\alpha_{1},\beta,{b}_{2},{b}_{3})\\
& -[{r}_{3}\phi^{a}_{T}({x}_{3})\phi^{T}_{T}({x}_{2})+{r}_{3}\phi^{v}_{T}({x}_{3})\phi^{T}_{T}({x}_{2})]{E}_{af}({t}_{f}){h}_{af}(\alpha_{2},\beta,{b}_{3},{b}_{2}) \}\\
\end{split}
\end{equation}

\begin{equation}
\begin{split}
{\emph F}^{\emph{LL,T}}_{af}=&4\pi\emph{C}_{\emph{F}}f_{B}M^{4}_{B^0_{s}}{r}_{2}{r}_{3}\int^{1}_{0}\emph{d}\emph{x}_{2}\emph{d}\emph{x}_{3}\int^\frac{1}{\Lambda}_{0}\emph{b}_{2}\emph{d}\emph{b}_{2}\emph{b}_{3}\emph{d}\emph{b}_{3}\phi_{B_{s}}(\emph{x}_{1},b_{1})\\
& \times \{[({x}_{3}-2)(\phi^{a}_{T}({x}_{2})\phi^{v}_{T}({x}_{3})+ \phi^{v}_{T}({x}_{2})\phi^{a}_{T}({x}_{3}))-{x}_{3}\phi^{v}_{T}({x}_{2})\phi^{a}_{T}({x}_{3})\\
&-{x}_{3}\phi^{a}_{T}({x}_{2})\phi^{v}_{T}({x}_{3})]{E}_{af}({t}_{e}){h}_{af}(\alpha_{1},\beta,{b}_{2},{b}_{3})\\
&+[({x}_{2}-1)\phi^{v}_{T}({x}_{3})\phi^{v}_{T}({x}_{2})+({x}_{2}-1)\phi^{a}_{T}({x}_{3})\phi^{a}_{T}({x}_{2})\\
&+({x}_{2}+1)\phi^{a}_{T}({x}_{3})\phi^{v}_{T}({x}_{2})+({x}_{2}+1)\phi^{v}_{T}({x}_{3})\phi^{a}_{T}({x}_{2})]{E}_{af}({t}_{f}){h}_{af}(\alpha_{2},\beta,{b}_{3},{b}_{2}) \}\\
\end{split}
\end{equation}
\begin{equation}
\begin{split}
{\emph F}^{\emph{SP,N}}_{af}=-{\emph F}^{\emph{SP,T}}_{af}
\end{split}
\end{equation}
\begin{equation}
\begin{split}
{\emph F}^{\emph{LL,T}}_{af}=-{\emph F}^{\emph{LR,T}}_{af}
\end{split}
\end{equation}

The transverse polarization amplitudes of the nonfactorizable annihilation diagrams are
\begin{equation}
\begin{split}
{\emph M}^{\emph{LL(SP),N}}_{anf}=&8\sqrt\frac{2}{3}\pi\emph{C}_{\emph{F}}M^{4}_{B^0_{s}}\int^{1}_{0}\emph{d}\emph{x}_{1}\emph{d}\emph{x}_{2}\emph{d}\emph{x}_{3}\int^\frac{1}{\Lambda}_{0}\emph{b}_{1}\emph{d}\emph{b}_{1}\emph{b}_{2}\emph{d}\emph{b}_{2}\phi_{B_{s}}(\emph{x}_{1},b_{1})\\
& \times \{[-2{r}_{2}{r}_{3} \phi^{a}_{T}({x}_{2})\phi^{a}_{T}({x}_{3})-2{r}_{2}{r}_{3}\phi^{v}_{T}({x}_{2}) \phi^{v}_{T}({x}_{3})\\
&-{r}^{2}_{2}(x_{2}-1)\phi^{T}_{T}({x}_{2})\phi^{T}_{T}({x}_{3})+{r}^{2}_{3}{x}_{3}\phi^{T}_{T}({x}_{2}) \phi^{T}_{T}({x}_{3})]{E}_{anf}({t}_{g}){h}_{anf}(\alpha,\beta_{1},{b}_{1},{b}_{2})\\
& +[{r}^{2}_{2}x_{2}\phi^{T}_{T}({x}_{2})\phi^{T}_{T}({x}_{3})-{r}^{2}_{3}(x_{3}-1)\phi^{T}_{T}({x}_{2})\phi^{T}_{T}({x}_{3})]{E}_{anf}({t}_{h}){h}_{anf}(\alpha,\beta_{2},{b}_{1},{b}_{2}) \}\\
\end{split}
\end{equation}

\begin{equation}
\begin{split}
{\emph M}^{\emph{LR,N}}_{anf}=&8\sqrt\frac{2}{3}\pi\emph{C}_{\emph{F}}M^{4}_{B^0_{s}}\int^{1}_{0}\emph{d}\emph{x}_{1}\emph{d}\emph{x}_{2}\emph{d}\emph{x}_{3}\int^\frac{1}{\Lambda}_{0}\emph{b}_{1}\emph{d}\emph{b}_{1}\emph{b}_{2}\emph{d}\emph{b}_{2}\phi_{B_{s}}(\emph{x}_{1},b_{1})\\
& \times \{[{r}_{2}({x}_{2}-2)\phi^{a}_{T}({x}_{2})\phi^{T}_{T}({x}_{3})+ {r}_{2}({x}_{2}-2)\phi^{v}_{T}({x}_{2})\phi^{T}_{T}({x}_{3})\\
&-{r}_{3}({x}_{3}+1)\phi^{T}_{T}({x}_{2})\phi^{a}_{T}({x}_{2})+{r}_{3}({x}_{3}+1)\phi^{T}_{T}({x}_{2})\phi^{v}_{T}({x}_{3})]{E}_{anf}({t}_{g}){h}_{anf}(\alpha,\beta_{1},{b}_{1},{b}_{2})\\
&+[{r}_{3}({x}_{3}-1)\phi^{T}_{T}({x}_{2})\phi^{v}_{T}({x}_{3})-{r}_{3}({x}_{3}-1)\phi^{T}_{T}({x}_{2})\phi^{a}_{T}({x}_{3})+{r}_{2}{x}_{2}\phi^{T}_{T}({x}_{3})\phi^{a}_{T}({x}_{2})\\
&+{r}_{2}{x}_{2}\phi^{T}_{T}({x}_{3})\phi^{v}_{T}({x}_{2})]{E}_{anf}({t}_{h}){h}_{anf}(\alpha,\beta_{2},{b}_{1},{b}_{2}) \}\\
\end{split}
\end{equation}
\begin{equation}
\begin{split}
{\emph M}^{\emph{LR,N}}_{anf}= -{\emph M}^{\emph{LR,T}}_{anf}\\
\end{split}
\end{equation}

\begin{equation}
\begin{split}
{\emph M}^{\emph{LL,T}}_{anf}=&8\sqrt\frac{2}{3}\pi\emph{C}_{\emph{F}}M^{4}_{B^0_{s}}\int^{1}_{0}\emph{d}\emph{x}_{1}\emph{d}\emph{x}_{2}\emph{d}\emph{x}_{3}\int^\frac{1}{\Lambda}_{0}\emph{b}_{1}\emph{d}\emph{b}_{1}\emph{b}_{2}\emph{d}\emph{b}_{2}\phi_{B_{s}}(\emph{x}_{1},b_{1})\\
& \times \{[2{r}_{2}{r}_{3} \phi^{a}_{T}({x}_{2})\phi^{v}_{T}({x}_{3})+2{r}_{2}{r}_{3}\phi^{v}_{T}({x}_{2}) \phi^{a}_{T}({x}_{3})\\
&-{r}^{2}_{2}(x_{2}-1)\phi^{T}_{T}({x}_{2})\phi^{T}_{T}({x}_{3})-{r}^{2}_{3}{x}_{3}\phi^{T}_{T}({x}_{2}) \phi^{T}_{T}({x}_{3})]{E}_{anf}({t}_{g}){h}_{anf}(\alpha,\beta_{1},{b}_{1},{b}_{2})\\
& +[{r}^{2}_{2}x_{2}\phi^{T}_{T}({x}_{2})\phi^{T}_{T}({x}_{3})+{r}^{2}_{3}(x_{3}-1)\phi^{T}_{T}({x}_{2})\phi^{T}_{T}({x}_{3})]{E}_{anf}({t}_{h}){h}_{anf}(\alpha,\beta_{2},{b}_{1},{b}_{2}) \}\\
\end{split}
\end{equation}
\begin{equation}
\begin{split}
{\emph M}^{\emph{LL,T}}_{anf}=-{\emph M}^{\emph{SP,T}}_{anf}
\end{split}
\end{equation}

For $B^{0}_{s} \rightarrow \ TT $ decays, the transverse polarization amplitudes of the nonfactorizable emission diagrams are as follows
\begin{equation}
\begin{split}
{\emph M}^{\emph{LL,N}}_{enf}=&8\sqrt\frac{2}{3}\pi\emph{C}_{\emph{F}}M^{4}_{B^0_{s}}{r}_{2}\int^{1}_{0}\emph{d}\emph{x}_{1}\emph{d}\emph{x}_{2}\emph{d}\emph{x}_{3}\int^\frac{1}{\Lambda}_{0}\emph{b}_{1}\emph{d}\emph{b}_{1}\emph{b}_{2}\emph{d}\emph{b}_{2}\phi_{B_{s}}(\emph{x}_{1},b_{1})\\
& \times \{[(1-{x}_{2}) \phi^{a}_{T}({x}_{2})\phi^{T}_{T}({x}_{3})+ (1-{x}_{2})\phi^{T}_{T}({x}_{3})\phi^{v}_{T}({x}_{2})]{E}_{enf}({t}_{c}){h}_{enf}(\alpha,\beta_{1},{b}_{1},{b}_{2})\\
&-[2{r}_{3}(x_{2}+x_{3})\phi^{a}_{T}({x}_{2})\phi^{a}_{T}({x}_{3})+2{r}_{3}(x_{2}+x_{3})\phi^{v}_{T}({x}_{2})\phi^{v}_{T}({x}_{3})-{x}_{2}(\phi^{T}_{T}({x}_{3})\phi^{a}_{T}({x}_{2})\\
&+\phi^{T}_{T}({x}_{3})\phi^{v}_{T}({x}_{2}))]{E}_{enf}({t}_{d}){h}_{enf}(\alpha,\beta_{2},{b}_{1},{b}_{2}) \}\\
\end{split}
\end{equation}
\begin{equation}
\begin{split}
{\emph M}^{\emph{LR,N}}_{enf}=&8\sqrt\frac{2}{3}\pi\emph{C}_{\emph{F}}M^{4}_{B^{0}_{s}}\int^{1}_{0}\emph{d}\emph{x}_{1}\emph{d}\emph{x}_{2}\emph{d}\emph{x}_{3}\int^\frac{1}{\Lambda}_{0}\emph{b}_{1}\emph{d}\emph{b}_{1}\emph{b}_{2}\emph{d}\emph{b}_{2}\phi_{B_{s}}(\emph{x}_{1},b_{1})\phi^{T}_{T}({x}_{2})\\
& \times \{[{r}_{3}{x}_{3} \phi^{a}_{T}({x}_{3})- {r}_{3}{x}_{3}\phi^{v}_{T}({x}_{3})-{r}^{2}_{2}({x}_{2}-1)\phi^{T}_{T}({x}_{3})+{x}_{3}{r}^{2}_{3}\phi^{T}_{T}({x}_{3})]{E}_{enf}({t}_{c}){h}_{enf}(\alpha,\beta_{1},{b}_{1},{b}_{2})\\
&+[{r}_{3}{x}_{3} \phi^{a}_{T}({x}_{3})- {r}_{3}{x}_{3}\phi^{v}_{T}({x}_{3})+{r}^{2}_{2}{x}_{2}\phi^{T}_{T}({x}_{3})+{x}_{3}{r}^{2}_{3}\phi^{T}_{T}({x}_{3})]{E}_{enf}({t}_{d}){h}_{enf}(\alpha,\beta_{2},{b}_{1},{b}_{2}) \}\\
\end{split}
\end{equation}
\begin{equation}
\begin{split}
{\emph M}^{\emph{SP,N}}_{enf}=&8\sqrt\frac{2}{3}\pi\emph{C}_{\emph{F}}M^{4}_{B^{0}_{s}}{r}_{2}\int^{1}_{0}\emph{d}\emph{x}_{1}\emph{d}\emph{x}_{2}\emph{d}\emph{x}_{3}\int^\frac{1}{\Lambda}_{0}\emph{b}_{1}\emph{d}\emph{b}_{1}\emph{b}_{2}\emph{d}\emph{b}_{2}\phi_{B_{s}}(\emph{x}_{1},b_{1})\\
& \times \{[2{r}_{3}({x}_{3}-{x}_{2}+1) \phi^{v}_{T}({x}_{2})\phi^{v}_{T}({x}_{3})- 2{r}_{3}({x}_{3}+1-{x}_{2})\phi^{a}_{T}({x}_{2})\phi^{a}_{T}({x}_{3})\\
&+({x}_{2}-1)\phi^{T}_{T}({x}_{3})\phi^{v}_{T}({x}_{2})-({x}_{2}-1)\phi^{T}_{T}({x}_{3})\phi^{a}_{T}({x}_{2})]{E}_{enf}({t}_{c}){h}_{enf}(\alpha,\beta_{1},{b}_{1},{b}_{2})\\
&+[{x}_{2} \phi^{T}_{T}({x}_{3})\phi^{a}_{T}({x}_{2})- {x}_{2}\phi^{v}_{T}({x}_{2})\phi^{T}_{T}({x}_{3})]{E}_{enf}({t}_{d}){h}_{enf}(\alpha,\beta_{2},{b}_{1},{b}_{2}) \}\\
\end{split}
\end{equation}
\begin{equation}
\begin{split}
{\emph M}^{\emph{LL,T}}_{enf}=&8\sqrt\frac{2}{3}\pi\emph{C}_{\emph{F}}M^{4}_{B^0_{s}}{r}_{2}\int^{1}_{0}\emph{d}\emph{x}_{1}\emph{d}\emph{x}_{2}\emph{d}\emph{x}_{3}\int^\frac{1}{\Lambda}_{0}\emph{b}_{1}\emph{d}\emph{b}_{1}\emph{b}_{2}\emph{d}\emph{b}_{2}\phi_{B_{s}}(\emph{x}_{1},b_{1})\\
& \times \{[({x}_{2}-1) \phi^{a}_{T}({x}_{2})\phi^{T}_{T}({x}_{3})+ ({x}_{2}-1)\phi^{T}_{T}({x}_{2})\phi^{v}_{T}({x}_{3})]{E}_{enf}({t}_{c}){h}_{enf}(\alpha,\beta_{1},{b}_{1},{b}_{2})\\
&+[2{r}_{3}(x_{2}+x_{3})\phi^{a}_{T}({x}_{2})\phi^{v}_{T}({x}_{3})+2{r}_{3}(x_{2}+x_{3})\phi^{v}_{T}({x}_{2})\phi^{a}_{T}({x}_{3})-{x}_{2}\phi^{T}_{T}({x}_{3})\phi^{a}_{T}({x}_{2})\\
&-x_{2}\phi^{T}_{T}({x}_{3})\phi^{v}_{T}({x}_{2})]{E}_{enf}({t}_{d}){h}_{enf}(\alpha,\beta_{2},{b}_{1},{b}_{2}) \}\\
\end{split}
\end{equation}

\begin{equation}
\begin{split}
{\emph M}^{\emph{LR,T}}_{enf}=&8\sqrt\frac{2}{3}\pi\emph{C}_{\emph{F}}M^{4}_{B^0_{s}}\int^{1}_{0}\emph{d}\emph{x}_{1}\emph{d}\emph{x}_{2}\emph{d}\emph{x}_{3}\int^\frac{1}{\Lambda}_{0}\emph{b}_{1}\emph{d}\emph{b}_{1}\emph{b}_{2}\emph{d}\emph{b}_{2}\phi_{B_{s}}(\emph{x}_{1},b_{1})\phi^{T}_{T}({x}_{2})\\
& \times \{[{r}_{3}{x}_{3} \phi^{v}_{T}({x}_{3})- {r}_{3}{x}_{3}\phi^{a}_{T}({x}_{3})-{r}^{2}_{2}({x}_{2}-1)\phi^{T}_{T}({x}_{3})-{x}_{3}{r}^{2}_{3}\phi^{T}_{T}({x}_{3})]{E}_{enf}({t}_{c}){h}_{enf}(\alpha,\beta_{1},{b}_{1},{b}_{2})\\
&+[{r}_{3}{x}_{3} \phi^{v}_{T}({x}_{3})- {r}_{3}{x}_{3}\phi^{a}_{T}({x}_{3})+{r}^{2}_{2}{x}_{2}\phi^{T}_{T}({x}_{3})-{x}_{3}{r}^{2}_{3}\phi^{T}_{T}({x}_{3})]{E}_{enf}({t}_{d}){h}_{enf}(\alpha,\beta_{2},{b}_{1},{b}_{2}) \}
\end{split}
\end{equation}

\begin{equation}
\begin{split}\label{equ:h}
{\emph M}^{\emph{SP,T}}_{enf}=&8\sqrt\frac{2}{3}\pi\emph{C}_{\emph{F}}M^{4}_{B^0_{s}}{r}_{2}\int^{1}_{0}\emph{d}\emph{x}_{1}\emph{d}\emph{x}_{2}\emph{d}\emph{x}_{3}\int^\frac{1}{\Lambda}_{0}\emph{b}_{1}\emph{d}\emph{b}_{1}\emph{b}_{2}\emph{d}\emph{b}_{2}\phi_{B_{s}}(\emph{x}_{1},b_{1})\\
& \times \{[2{r}_{3}({x}_{2}-{x}_{3}-1) \phi^{v}_{T}({x}_{2})\phi^{a}_{T}({x}_{3})- 2{r}_{3}({x}_{3}+1-{x}_{2})\phi^{a}_{T}({x}_{2})\phi^{v}_{T}({x}_{3})\\
&+({x}_{2}-1)\phi^{T}_{T}({x}_{3})\phi^{a}_{T}({x}_{2})-({x}_{2}-1)\phi^{T}_{T}({x}_{3})\phi^{v}_{T}({x}_{2})]{E}_{enf}({t}_{c}){h}_{enf}(\alpha,\beta_{1},{b}_{1},{b}_{2})\\
&+[{x}_{2} \phi^{T}_{T}({x}_{3})\phi^{v}_{T}({x}_{2})- {x}_{2}\phi^{a}_{T}({x}_{2})\phi^{T}_{T}({x}_{3})]{E}_{enf}({t}_{d}){h}_{enf}(\alpha,\beta_{2},{b}_{1},{b}_{2}) \}\\
\end{split}
\end{equation}
\section{Numerical results and discussions}\label{sec:numer}
In this section, we initiate our calculations by enumerating the input parameters. These encompass the decay constant $f_{B}$, the Wolfenstein parameters, the masses of $B$ meson and tensor mesons, and the corresponding lifetime, as detailed in Table~\ref{tab1}~\cite{Particle Date Group:2020ac}.
\begin{table}[htbp]
	\centering
	\caption{Various parameters involved in the calculation process}
	\label{tab1}
	\begin{tabular*}{\columnwidth}{@{\extracolsep{\fill}}lllll@{}}
		\hline
		\hline
		\\
		Mass of mesons   &$M_{B^{0}_{s}}=5.367$ {\rm GeV}     &$M_{a_{2}(1320)}=1.317$ {\rm GeV}       &$M_{K^{*0}_{2}(1430)}=1.432$ {\rm GeV} \\
		\\		
		 &$M_{K^{*\pm}_{2}(1430)}=1.427$ {\rm GeV}	&$M_{f^{'}_{2}(1525)}=1.517$ {\rm GeV} &$M_{f_{2}(1270)}=1.275$ {\rm GeV}\\
		\\
		&$m_{b}=4.18$ {\rm GeV}              &$m_{s}=0.093$ {\rm GeV}       \\
		\\
		Decay constants of mesons           &$f_{B^0_{s}}=0.24 \pm 0.02$ {\rm GeV}  \\
		\\
		Lifetime of meson               &$\tau_{B^{0}_{s}}=1.509$ {\rm ps}     \\
		\\
		Wolfenstein parameters           &$\emph{A}=0.836\pm0.015$              &$\lambda=0.22453\pm0.00044$
		\\
		&$\bar{\eta}=0.355^{+0.012}_{-0.011}$
		&$\bar{\rho}=0.122^{+0.018}_{-0.017}$\\
		\\
		\hline
		\hline
	\end{tabular*}
\end{table}

Our numerical calculations within the pQCD framework are focused on branching ratios, direct $CP$ violations, and polarization fractions, as summarized in Tables~\ref{2223}-\ref{7777}.  It is crucial to acknowledge that there exist uncertainties in our calculation results. In Table~\ref{2223}, the errors stem from induced by the uncertainties in the  the shape parameter $\omega_{B}=(0.50\pm 0.05)$ GeV pertaining to the $B^{0}_{s} $ meson distribution amplitude~\cite{Ali:2007ff}. The second source of uncertainty pertains to the $B^{0}_{s} $ meson and the final state tensor mesons, as documented in Table~\ref{111}. The third error arises from $\Lambda_{QCD}$= $ (0.25\pm{0.05}) $ GeV, and varies $ 20\% $ from hard scale $t_{max}=(1.0\pm 0.2)t$ detailed in the Appendix. Other uncertainties such as the Cabibbo-Kobayashi-Maskawa matrix elements V from the $\bar{\eta}$ and $\bar{\rho}$, angles of the unitary triangle that can be neglected.

With the amplitudes calculated in  Sec.~\ref{sec:amp}, the decay width is determined as
\begin{equation}
\begin{split}
\Gamma={\frac{[(1-(r_{2}+r_{3})^{2})(1-(r_{2}-r_{3})^{2})]^{1/2}}{16 \pi m_{B}}}\sum_{i}\mid A_{i}\mid^{2}.
\end{split}
\end{equation}

The branching ratio is got through $  \cal {BR}$ = $\Gamma\cdot \tau_{B^{0}_{s}}$. In Ref.~\cite{Bigi:2000yz,Branco:1999fs}, we can learn about direct $CP$ violations, $ A^{dir}_{CP} $ is defined by
\begin{equation}
\begin{split}
{A^{dir}_{CP}}={\frac{\mid \overline A_{\overline f}\mid^{2}-\mid A_{f}\mid^{2}}{\mid\overline A_{\overline f}\mid^{2}+\mid A_{ f}\mid^{2}}}.\\
\end{split}
\end{equation}

Here the two amplitudes are defined as follows
\begin{equation}
\begin{split}
 A_{f}= < f\mid\cal H\mid B> ,\\
\overline A_{\overline {f}} =< \bar{f} \mid\cal H\mid \bar{B} >,
\end{split}
\end{equation}
where the $\overline B$ meson has a $ b $ quark and $ \overline f $ is the $CP$ conjugate state of $ f $. The results of the polarization fractions $ f_{i} $, which are defined as
\begin{equation}
\begin{split}
{f_{0,\|,\perp}}={\frac{\mid A_{i}\mid^{2}}{\sum_{i}\mid A_{i}\mid^{2}}},
\end{split}
\end{equation}
where $ A_{i}(i=0,\|,\perp) $ is the amplitude of the longitudinal or transverse polarization contributions. Based on the helicity amplitudes $A_{i} (i=0,\|,\perp)$ for longitudinal, parallel, and perpendicular polarizations, the three part amplitudes  are given as
\begin{equation}
\begin{split}
A_{0}=\xi m^{2}_{B^{0}_{s}}A_{L},
A_{\|}=\xi\sqrt{2} m ^{2}_{B^{0}_{s}}A_{N},\\
A_{\perp}=\xi\sqrt{2(r^{2}-1)}
m ^{2}_{B^{0}_{s}}r_{2}r_{3}A_{T},
\end{split}
\end{equation}
here $\xi=\sqrt{\frac{G^{2}_{F}P_{c}}{(16\pi m^{2}_{B}\Gamma)}}$ , and the ratio $ r=\frac{P_{2}\cdot P_{3}}{(m^{2}_{B}r_{2}r_{3})} $.\\

\begin{table}[htbp]
	\centering
	\caption{The $CP$-averaged branching ratios of the $B^{0}_{s} \rightarrow T_{1}T_{2} $ decay (in unit of $10^{-6}$), the errors  attributed to the shape parameter, decay constants, hard scale and QCD scale.}
	\label{2223}
	\begin{tabular*}{\columnwidth}{@{\extracolsep{\fill}}lllll@{}}
		\hline
		\hline
		Decay Modes 	&$B_{0}$ &$B_{\parallel}$ &$B_{\perp}$  &$B_{total}$ \\
		\hline
		\\
		$B^{0}_{s}\rightarrow a^{0}_{2}a^{0}_{2}$   &$0.28^{+0.01+0.08+0.18}_{-0.03-0.02-0.01}$
	&$0.03^{+0.00+0.01+0.00}_{-0.00-0.00-0.00}$  &$0.00$  &$0.31^{+0.01+0.09+0.18}_{-0.03-0.02-0.01}$      \\
		\\
		$B^{0}_{s}\rightarrow  a^{+}_{2}a^{-}_{2}$  &$0.42^{+0.01+0.04+0.32}_{-0.03-0.02-0.01}$
		&$0.06^{+0.01+0.02+0.00}_{-0.01-0.01-0.02}$    &$0.00$   &$0.48^{+0.02+0.06+0.32}_{-0.04-0.03-0.03}$ \\
		\\
		$B^{0}_{s}\rightarrow\ K^{*0}_{2}\overline{\emph{K}}^{*0}_{2}$
		&$2.03^{+1.74+0.91+1.45}_{-1.64-0.65-1.20}$
	 &$0.22^{+0.17+0.08+0.10}_{-0.10-0.07-0.06}$   &$0.47^{+0.33+0.20+0.24}_{-0.18-0.14-0.20}$ &$2.72^{+2.24+1.18+1.79}_{-1.92-0.86-1.46}$ \\
		\\
		$B^{0}_{s}\rightarrow\ K^{*+}_{2}{K}^{*-}_{2}$
	&$1.86^{+1.62+0.83+1.34}_{-1.40-0.59-1.11}$
	&$0.21^{+0.16+0.08+0.09}_{-0.09-0.06-0.01}$ &$0.19^{+0.13+0.08+0.17}_{-0.08-0.06-0.08}$ &$2.26^{+1.91+0.99+1.60}_{-1.57-0.71-1.20}$ \\
		\\
		$B^{0}_{s}\rightarrow\ f_{2}f_{2}$ &$0.50^{+0.04+0.08+0.24}_{-0.05-0.04-0.01}$   &$0.04^{+0.00+0.01+0.00}_{-0.00-0.00-0.00}$   &$0.00$
		&$0.54^{+0.04+0.09+0.24}_{-0.05-0.04-0.01}$ \\
		\\
		$B^{0}_{s}\rightarrow\ f_{2}f^{'}_{2}$&$1.86^{+1.13+0.76+1.68}_{-0.65-0.57-0.87}$
	&$0.16^{+0.14+0.07+0.17}_{-0.07-0.05-0.01}$   &$0.44^{+0.31+0.08+0.42}_{-0.17-0.07-0.30}$ &$2.46^{+1.58+0.91+2.27}_{-0.89-0.69-1.18}$ \\
		\\
		$B^{0}_{s}\rightarrow\ f^{'}_{2}f^{'}_{2}$ &$6.03^{+0.84+1.68+4.51}_{-2.38-1.86-3.49}$        &$0.42^{+0.38+0.14+0.01}_{-0.21-0.12-0.02}$  &$1.23^{+1.23+0.47+0.01}_{-0.65-0.38-0.16}$ &$7.68^{+2.45+2.29+4.53}_{-3.24-2.36-3.67}$\\
		\\
		\hline
		\hline
	\end{tabular*}
\end{table}

Table~\ref{2223} displays the pertinent data. Several observations can be made:

(1) For $B^{0}_{s}\rightarrow\ a^{0}_{2}a^{0}_{2} , a^{+}_{2}a^{-}_{2}, f_{2}f_{2}$, they are only the pure annihilation diagrams, whose branching ratios are at the order of $ 10^{-7} $. For $B^{0}_{s}\rightarrow\ f^{'}_{2}f^{'}_{2}, f_{2}f^{'}_{2}, K^{*0}_{2}\overline{\emph{K}}^{*0}_{2},K^{*+}_{2}{\emph{K}}^{*-}_{2}$, they have the annihilation and emission diagrams, whose branching ratios are at the order of $ 10^{-6} $. Under the $ SU(3) $ limit, since the Bose statistics are satisfied, the meson wave function will be antisymmetric when the momenta fractions of the quark and antiquark of tensor mesons are exchanged~\cite{Cheng:2010hn,Cheng:2010yd}. Due to the commutative antisymmetry of the tensor meson wave function, the nonfactorizable emission diagrams will be more pronounced  and  provide a greater contribution~\cite{Cheng:2010hn,Cheng:2010yd}. Take $B^{0}_{s}\rightarrow\ f^{'}_{2}f^{'}_{2}$ for example, if without nonfactorizable emission contributions, the branching ratio will decrease $ 90\% $, and its longitudinal polarization fraction will also reduce largely, which is different from the $B^{0}_{s}\rightarrow\ f_{1}(1420)f_{1}(1420)$ in Ref.\cite{Jiang:2020eml}. For $B^{0}_{s}\rightarrow\ f_{1}(1420)f_{1}(1420)$, the annihilation diagrams could contribute a large imaginary part and play an important role in calculating the branching ratios.
The reason for the difference may be that, relative to the commutative antisymmetry of the tensor meson wave function, nonfactorizable emission contributions do not get offset but enhanced.

(2) From the Ref.~\cite{Kim:2013cpa}, we find that $ {\cal BR}(B^{0}_{s}\rightarrow\ \phi (K^{*-}_{2}(1430),\overline{\emph{K}}^{*0}_{2}(1430))) $ and $ {\cal BR}(B^{0}_{s}\rightarrow\ \phi (K^{*-}(1430) \overline{\emph{K}}^{*0}(1430)))$ are at the same order, but the former is a little small. The authors observed that only small effects when $ {K}^{*0}(1430) $ is substituted by $ K^{*}_{2}(1430) $. We observe  the corresponding decays in Ref.~\cite{PDG}, such as $B^{0}_{s}\rightarrow\ K^{*\pm}_{0}(1430){\emph{K}}^{\pm} $ and  $B^{0}_{s}\rightarrow\ K^{*}_{0}(1430)\overline{\emph{K}}^{0} $, the branching ratio is also at the order of $ 10^{-5} $ when vector mesons are replaced by tensor mesons. The branching ratio of $B^{0}_{s}\rightarrow\ K^{*0}(1430)\overline{\emph{K}}^{*0}_{2}(1430) $ or $B^{0}_{s}\rightarrow\ K^{*0}_{2}(1430)\overline{\emph{K}}^{*0}(1430)$ is at the order of $ 6\times 10^{-6}\sim 9\times 10^{-6} $, and the branching ratio of $B^{0}_{s}\rightarrow\ K^{*0}_{2}(1430)\overline{\emph{K}}^{*0}_{2}(1430)$ is at the order of $ 2.71\times10^{-6} $ in this paper, which also supports the view of the authors of Ref.~\cite{Kim:2013cpa}.

(3) For $B^{0}_{s}\rightarrow\ VT(a_{2},f_{2})$, when a vector meson is emitted, the factorizable emission contribution of the penguin diagrams will offset the contribution of the tree annihilation diagrams, and the branching ratio turn to be very small. However, the contribution of the annihilation diagrams does not get offset due to the absence of factorizable emission diagrams in $B^{0}_{s}\rightarrow\ TT$. Therefore, the branching ratio of $B^{0}_{s}\rightarrow\ TT(a_{2},f_{2})$ is one or two orders of magnitude larger than that of $B^{0}_{s}\rightarrow\ VT(a_{2},f_{2})$, making it more beneficial to experimental observation.

\begin{figure}[htbp]
	\centering
	\begin{tabular}{l}
		\includegraphics[width=0.5\textwidth]{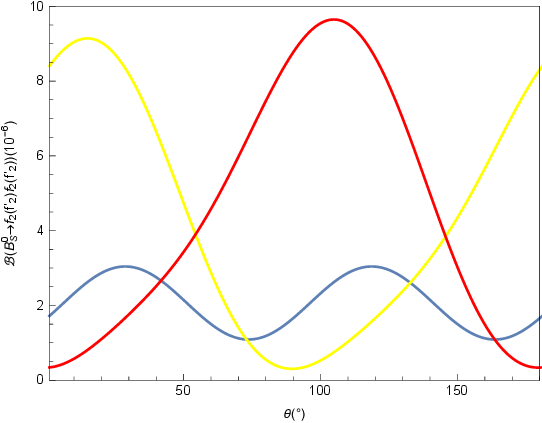}
	\end{tabular}
	\caption {The branching ratio of ${B}^{0}_{s} \rightarrow f_{2}(f^{'}_{2})f_{2}(f^{'}_{2})$ decay on the mixing angle $\theta$, in which the red, the yellow, and the blue solid lines correspond to the $B^{0}_{s}$ decays with final states $f_{2}(1270)f_{2}(1270) $,$f^{'}_{2}(1525)f^{'}_{2}(1525)$, and $f_{2}(1270)f^{'}_{2}(1525)$, respectively.}
	\label{fig2}
\end{figure}

(4) For $B^{0}_{s}\rightarrow\ f_{2}(f^{'}_{2})f_{2}(f^{'}_{2})$ decays with $ f_{2}(1270)-f^{'}_{2}(1525) $ mixing, just as the $ \eta-\eta^{'} $ mixing. To see the variation clearly with the mixing angle, we show the branching ratios  $ \cal {B}$ $(B^{0}_{s}\rightarrow\ f_{2}(f^{'}_{2})f_{2}(f^{'}_{2}))$ varying with $ \theta \in  [0,\pi]$ in Fig.~\ref{fig2}. In Ref.\cite{Jiang:2020eml}, the authors plotted the related figures about the branching ratios of $B^{0}_{s}\rightarrow\ f_{1}f_{1}$ decays dependent on the free parameter $\theta$. By comparing  figures about $B^{0}_{s}\rightarrow\ f_{1}f_{1}$ and $B^{0}_{s}\rightarrow\ f_{2}(f^{'}_{2})f_{2}(f^{'}_{2})$, we can find that when the $\theta$ is large enough, and its influence on the branching ratio is more obvious, which can be seen from  the line shapes.  When $\theta$ reaches a certain angle, the branching ratios of $B^{0}_{s}\rightarrow\ f_{2}(f^{'}_{2})f_{2}(f^{'}_{2})$ and  $B^{0}_{s}\rightarrow\ f_{1}f_{1}$ will vary an order of magnitude. But for $ f_{2}(1270)-f^{'}_{2}(1525) $ mixing, in the contrast to $ f_{1}(1285)-f_{1}(1420)$, the mixing angle is very small. From the Refs.~\cite{ParticleDataGroup:2008zun,Li:2000zb,Cheng:2011fk}, we have known that the mixing $ \theta $ is about $5.8 ^{\circ} \sim 10 ^{\circ} $, and the related branching ratios are close to that of $ 0 ^{\circ}$. In addition, the branching ratio of the $B^{0}_{s}\rightarrow\ f^{'}_{2}(1525)f^{'}_{2}(1525)$ decay is larger than that of the  $B^{0}_{s}\rightarrow\ f_{2}(1270)f_{2}(1270)$ decay by one order of magnitude, which is caused due to the reason that the former has more Feynman diagrams.

$B^{0}_{s}$ mesons can be produced in the $\Upsilon (5S)$ decays to $B^{0}_{s}\overline{\emph{B}}^{0}_{s}$,$B^{0}_{s}\overline{\emph{B}}^{*}_{s}$,$B^{*}_{s}\overline{\emph{B}}^{0}_{s}$ or $B^{*}_{s}\overline{\emph{B}}^{*}_{s}$  intermediate decay~\cite{Aushev:2010bq}. Since of the kinematic smearing from excited $B^{*}_{s}$ production and contamination from $B^{+/0}$ decays, measurements  are not as easy as at the  $\Upsilon (4S) $. Although full $B^{0}_{s}$ reconstruction would mitigate $B^{+/0}$ background, such a technique will only be possible with Belle II~\cite{Urquijo:2013npa}.
The number of $B^{0}_{s}$ mesons in a dataset can be calculated as: $N_{B^{0}_{s}} =2\times \sigma^{\Upsilon (5S)}_{b\overline{\emph{b}}}\times f_{s}\times\cal L$.  The parameter  $f_{s}$  is a key component to calculate the total $B^{0}_{s}$ yield in the sample. $\cal L$ is the integrated luminosity of the data and $\sigma^{\Upsilon (5S)}_{b\overline{\emph{b}}}$ is the cross section of the process $e^{+}e^{-}\rightarrow  b\overline{\emph{b}}$ ~\cite{Aushev:2010bq,Oswald:2013tna}. By using the values from the Refs.~\cite{CLEO:2006dkk,Belle:2006jvm}, the number of $B^{0}_{s}$ mesons is estimated to be $\sim5.9\times10^{8}$ in the dataset of $\cal L$ = $ 5 ab^{-1}$ taken at  $\Upsilon (5S)$ in Belle II ~\cite{Aushev:2010bq}, which indicates that a $ 5 ab^{-1}\Upsilon (5S) $ sample contains approximately 300 million $B^{0}_{s}\overline{\emph{B}}^{0}_{s}$ pairs. The branching ratios of our calculation for $B^{0}_{s} \rightarrow \ TT $ are at the order of $10^{-6}$ and $10^{-7}$. Therefore,  the decays of $B^{0}_{s} \rightarrow \ TT $ will hopefully be observed by the Belle II experiments in the near future.

\begin{table}[htbp]
	\centering
	\caption{Presents the polarization fraction of the decay. The accompanying errors arise from considerations encompassing the  shape parameter, decay constants, hard scale and QCD scale.}
	\label{a}
	\begin{tabular*}{\columnwidth}{@{\extracolsep{\fill}}lllll@{}}
		\hline
		\hline
		Decay Modes &$f_{0}$ &$f_{\parallel}$ &$f_{\perp}$ \\
		\hline
		\\
		$B^{0}_{s}\rightarrow a^{0}_{2}a^{0}_{2}$  &$90.92^{+0.11+0.15+3.58}_{-0.45-0.05-0.53}\%$
		&$8.88^{+0.46+0.06+0.56}_{-0.12-0.14-3.56}\%$ &$0.19\%$\\
		\\
		$B^{0}_{s}\rightarrow  a^{+}_{2}a^{-}_{2}$  &$88.24^{+0.29+0.65+6.83}_{-0.83-1.81-0.52}\%$
		 &$11.66^{+0.84+1.79+0.54}_{-0.30-0.65-6.82}\%$   &$0.09\%$   \\
		\\
		$B^{0}_{s}\rightarrow\ K^{*0}_{2}\overline{\emph{K}}^{*0}_{2}$
		&$75.02^{+1.22+0.43+2.40}_{-25.37-0.55-14.29}\%$
		&$8.08^{+7.58+0.14+12.27}_{-0.16-0.30-1.05}\%$   &$16.89^{+18.01+0.42+2.03}_{-1.05-0.13-1.38}\%$ \\
		\\
		$B^{0}_{s}\rightarrow\ K^{*+}_{2}{K}^{*-}_{2}$
		&$82.50^{+1.01+0.43+0.46}_{-15.80-0.38-11.02}\%$
		&$9.16^{+7.70+0.16+9.44}_{-0.41-0.35-1.46}\%$ &$8.34^{+8.10+0.22+0.99}_{-0.60-0.09-1.64}\%$\\
		\\
		$B^{0}_{s}\rightarrow\ f_{2}f_{2}$  &$93.7^{+0.01+0.77+2.35}_{-0.07-0.74-0.38}\%$ &$6.78^{+0.08+0.73+0.40}_{-0.02-0.75-2.33}\%$  &$0.14\%$\\
		\\
		$B^{0}_{s}\rightarrow\ f_{2}f^{'}_{2}$ &$75.78^{+1.44+2.22+2.70}_{-1.61-2.78-7.92}\%$
    	&$6.32^{+0.99+0.18+15.90}_{-0.88-0.30-3.18}\%$   &$17.91^{+0.60+3.07+0.47}_{-0.57-2.47-7.99}\%$ \\
    	\\
    	$B^{0}_{s}\rightarrow\ f^{'}_{2}f^{'}_{2}$ &$78.56^{+3.78+0.31+7.94}_{-10.88-1.28-15.12}\%$  &$5.44^{+2.45+0.19+4.42}_{-0.77-0.01-1.96}\%$ &$16.99^{+8.43+1.09+10.69}_{-3.02-0.06-5.99}\%$\\
		\\
		\hline
		\hline
	\end{tabular*}
\end{table}

Moving on to Table~\ref{a}, we delve into predictions pertaining to the polarization fraction of mesons. The pQCD approach has effectively elucidated the theoretical underpinnings of pure annihilation diagrams about
$ B^{0}_{s}\rightarrow\ \pi^{+}\pi^{-} $ and  $ B^{0}\rightarrow\ D^{-}_{S}K^{+} $ theoretically, and corresponding numerical results has been confirmed by experiments. Based on  this success, it is plausible to assert that the pQCD approach holds substantial predictive power for processes primarily governed by annihilation diagrams~\cite{Ali:2007ff,Lu:2002iv,ParticleDataGroup:2012pjm,Li:2008ts}. In line with prior research ~\cite{Cheng:2009mu,Zou:2015iwa} . it has been ascertained that the contributions to these processes are chiefly orchestrated by longitudinal polarization in the case of pure annihilation of two-body decays. The fractions pertaining to these decays have been observed to approach nearly $ 100\% $. This trend corroborates our predictions of $B^{0}_{s}\rightarrow\ f_{2}f_{2}$, $B^{0}_{s}\rightarrow\ a^{0}_{2}a^{0}_{2}$ and $B^{0}_{s}\rightarrow\ a^{+}_{2}a^{-}_{2}$. Notably, the longitudinal polarizations of these three decays approximate $ 90\% $, underscoring the indispensability of accounting for transverse polarization, which can yield noteworthy contributions in pure annihilation decays. To illustrate this point, we take  $ B^{0}_{s}\rightarrow\ a^{0}_{2}a^{0}_{2}$ and $ B^{0}_{s}\rightarrow\ f_{2}f_{2} $ as an example, from Table~\ref{9999} reveals that the contribution of longitudinal polarization surpasses that of transverse polarization, rendering the former the predominant factor.

Across all the decays calculated in this study, longitudinal polarization predominantly steers the main decay modes. Through a factorial power estimation, it is evident that longitudinal polarization will play a leading role in the  $ B $ meson decay~\cite{Kagan:2004uw,Korner:1979ci}.
For $B^{0}_{s}\rightarrow\ TT $, in the longitudinal polarization part, the contribution of nonfactorizable emission diagrams is substantial, and the extent of cancellation with annihilation diagrams is notably feeble. Conversely, in the transverse polarization segment, nonfactorizable emission diagrams contribute minimally, and this contribution effectively counterbalances the annihilation contribution. Consequently, longitudinal polarization prevails, and the transverse polarization fraction is about $ 10\% $-$ 30\% $.
\begin{table}[htbp]
	\centering
	\caption{The $CP$-violating asymmetries of the $B^{0}_{s} \rightarrow T_{1}T_{2} $ decay, the errors come from the shape parameter, decay constants, hard scale and QCD scale.}
	\label{7777}
	\begin{tabular*}{\columnwidth}{@{\extracolsep{\fill}}lllll@{}}
		\hline
		\hline
		Decay modes  &${\cal A}^{dir}_{CP}$  &${\cal A}^{dir}_{CP}(0)$  &${\cal A}^{dir}_{CP}(\parallel)$ &${\cal A}^{dir}_{CP}(\perp)$ \\
		\hline
		\\
		$B^{0}_{s}\rightarrow a^{0}_{2}a^{0}_{2}$   &$(3.30^{+3.34+11.88+13.19}_{-3.75-1.77-38.50})\%$        &$(3.39^{+4.05+15.04+17.03}_{-4.62-2.19-44.62})\%$     &$(2.21^{+0.61+0.13+0.31}_{-0.62-0.15-0.24})\%$   &$(13.80^{+0.88+1.23+1.10}_{-0.90-1.23-1.58})\%$              \\
		\\
		$B^{0}_{s}\rightarrow f_{2}f_{2}$ &$(-6.12^{+3.06+5.93+9.96}_{-3.00-3.45-33.23})\%$  &$(-6.73^{+3.25+6.34+10.71}_{-4.64-2.35-34.63})\%$    &$(1.80^{+0.48+0.14+0.25}_{-0.46-0.15-0.31})\%$          &$(9.14^{+0.64+0.89+1.87}_{-0.03-0.90-1.59})\%$            \\
		\\
		$B^{0}_{s}\rightarrow K^{*0}_{2}\overline{\emph{K}}^{*0}_{2}$ &$0.00\%$     &...      &...  &...\\
		\\
		$B^{0}_{s}\rightarrow f_{2}f^{'}_{2}$ &$(1.04^{+0.19+1.81+8.43}_{-0.68-0.35-4.58})\%$  &$(1.04^{+0.31+0.25+3.38}_{-0.55-0.29-7.51})\%$    &$(-71.35^{+2.18+0.10+19.17}_{-1.68-0.11-7.38})\%$          &$(26.72^{+6.84+2.83+18.02}_{-6.20-0.57-4.90})\%$     \\
		\\
		$B^{0}_{s}\rightarrow f^{'}_{2}f^{'}_{2}$ &0.00\%    &...    &...   &...  \\
		\\
		$B^{0}_{s}\rightarrow  a^{+}_{2}a^{-}_{2}$ &$(2.09^{+2.58+11.87+10.92}_{-3.03-1.46-22.78})\%$  &$(2.21^{+2.58+11.87+10.92}_{-3.03-1.46-22.78})\%$    &$(1.06^{+0.30+0.07+61.22}_{-0.29-0.07-0.12})\%$          &$(17.54^{+0.98+1.84+0.81}_{-1.02-1.79-1.84})\%$            \\
		\\
		$B^{0}_{s}\rightarrow  K^{*+}_{2}K^{*-}_{2}$ &($2.61^{+3.94+0.77+6.18}_{-36.51-0.70-5.13})\%$  &$(3.57^{+3.74+0.99+6.66}_{-53.07-0.99-9.60})\%$   &$(-2.44^{+0.05+0.12+3.51}_{-0.33-0.43-3.87})\%$
		&$(-0.91^{+6.62+0.70+2.48}_{-0.84-0.28-4.33})\%$            \\
				
		\hline
		\hline
	\end{tabular*}
\end{table}
\begin{table}[htbp]
	\centering
	\caption{Decay amplitudes (in unit of $10^{-3}~\rm GeV^{3}$)  of the $B^{0}_{s} \rightarrow T_{1}T_{2} $ modes with three polarizations in the pQCD approach, where only the central values are quoted for clarification.}
	\label{9999}
	\begin{tabular*}{\columnwidth}{@{\extracolsep{\fill}}lllll@{}}
		\hline
		\hline
		Components  &$B^{0}_{s}\rightarrow\ a^{0}_{2}a^{0}_{2}$ & $B^{0}_{s}\rightarrow\ a^{0}_{2}a^{0}_{2}$ &$B^{0}_{s}\rightarrow\ f_{2}f_{2}$ & $B^{0}_{s}\rightarrow\ f_{2}f_{2}$\\
		classification  & Tree diagrams &Penguin diagrams &Tree diagrams &Penguin diagrams\\
		\hline
		\\
		$A_{L}$           &$0.157-0.130i$          &$-0.082-0.224i$         &$-0.290-0.304i$  &$-0.353-0.615i$  \\
		\\
		$A_{N}$           &$-0.014+0.025i$         &$0.104+0.135i$          &$-0.015+0.030i$  &$0.130+0.148i$    \\
		\\
		$A_{T}$           &$0.010+0.017i$          &$-7.887\times10^{-3}+3.704\times10^{-3}i$             &$0.011+0.020i$          &$-9.728\times10^{-3}+3.298\times10^{-3}i$            \\
		
		\\
		\hline
		\hline
	\end{tabular*}
\end{table}

Direct $CP$ asymmetries of $B^{0}_{s} \rightarrow\ TT $ decays are listed in Table~\ref{7777}.
The magnitude of the direct $CP$ violation is proportional to the ratio of the penguin and tree contributions~\cite{ParticleDataGroup:2008zun}. For the $B^{0}_{s} \rightarrow\ VV $ and $B^{0}_{s} \rightarrow\ VT $, when penguin contributions and tree contributions stay at the same level, the direct $CP$ violation appears. Since the decays presented in this paper are dominated by the penguin contributions, so the direct $CP$ violation is very small. However, we can find that the direct $CP$ violations of some special channels in the transverse polarization are sizable, and the tree contributions become comparable, which brings relatively large direct $CP$ violation.

For pure annihilation-type decay, the $CP$-violating asymmetry is small, which has been pointed out in previous predictions of two-body decays~\cite{Zou:2015iwa,Cheng:2009mu}. From  Table~\ref{7777}, the $CP$-violating asymmetries about $B^{0}_{s}\rightarrow a^{0}_{2}a^{0}_{2}, f_{2}f_{2} $ and $ a^{+}_{2}a^{-}_{2}$, which also suggest that the $CP$-violating asymmetry of pure annihilation decay is very small. For $B^{0}_{s}\rightarrow\ K^{*0}_{2}(1430)\overline{\emph{K}}^{*0}_{2}(1430)$, $ f^{'}_{2}(1525)f^{'}_{2}(1525)$ (without considering the mixing angle), in the standard model, there is no contribution of the tree diagrams operator, so the direct $CP$ violation is zero.

\section{Summary} \label{sec:summary}
In this study, we predicted the relevant parameters of decays $B^{0}_{s} \rightarrow \ TT $ in the pQCD  approach, where the tensor mesons are $a_{2}(1320)$, $ f_{2}(1270) $, $ K^{*}_{2}(1430)$, and $ f^{'}_{2}(1525) $. We calculate the branching ratios, the polarization fractions, and the direct $CP$ violations of these decays. Our calculation results suggest that (1) the production of tensor mesons via either vector or tensor currents is prohibited, highlighting the significance of nonfactorizable emission and annihilation contributions. Notably, the nonfactorizable emission diagram exhibits an augmented contribution owing to the antisymmetry inherent in the tensor meson wave function. (2) For decays characterized exclusively by annihilation processes, the branching ratio is situated at an order of $10^{-7} $. And for $ f^{'}_{2}(1525) $, $ K^{*}_{2}(1430) $, they have a sizable branching ratio with the order of $ 10^{-6} $, which would be easy to be verified experimentally.
(3) Regarding polarization fractions, the preponderance of the decay processes investigated in this paper predominantly manifests longitudinal contributions, particularly in the case of pure annihilation decay, where it can reach around $90\% $. (4) The direct $CP$ violations  associated with these decays are of nominal magnitude. The interference between penguin and tree contributions will bring the direct   violations, and the two components assessed in this paper are not comparable.  Consequently, the direct $CP$ violations in the majority of channels remain negligible, with only select channels exhibiting relatively pronounced direct $CP$ violation appears.  In conclusion, we anticipate that our results can be scrutinized through forthcoming experiments at LHC-b and Belle II. Furthermore, we hope that these findings contribute positively to our understanding of the QCD behavior of tensor mesons.


\section*{ACKNOWLEDGMENTS}
The authors would to thank Hong Yang and Wen Liu for some valuable discussions.
This work is supported by the National Natural Science Foundation of China under Grant No. 11047028.
\section*{APPENDIX: FORMULAS FOR THE CALCULATION USED IN THE TEXT} \label{sec:appendix}
\appendix
\setcounter{equation}{0}
\renewcommand\theequation{A.\arabic{equation}}
In this section, we list the helicity amplitudes for every considered two-body decays of $ B $ meson.

\begin{equation}
\begin{split}
A_{h}(B^{0}_{s} \rightarrow a^{0}_{2}(1320)a^{0}_{2}(1320))
&=G_{F}({V^{*}_{ub}V_{us}[(C_{1}+\frac{1}{3}C_{2})F^{LL,h}_{a}+C_{2}M^{LL,h}_{a}]}\\
&-V^{*}_{tb}V_{ts}[({{2}C_{3}+\frac{2}{3}C_{4}+{2}C_{5}+\frac{2}{3}C_{6}}\\
&+{\frac{1}{2}C_{7}+\frac{1}{6}C_{8}+\frac{1}{2}C_{9}+\frac{1}{6}C_{10}})F^{LL,h}_{a}\\
&+({2}C_{4}+\frac{1}{2}C_{10})M^{LL,h}_{a}+({2}C_{6}+\frac{1}{2}C_{8})M^{SP,h}_{a}])
\end{split}
\end{equation}
and $B^{0}_{s}\rightarrow\ f_{2}(1270)f_{2}(1270) $  decay has the same amplitude as $ B^{0}_{s} \rightarrow a^{0}_{2}(1320)a^{0}_{2}(1320) $ decays.

\begin{equation}
\begin{split}
A_{h}(B^{0}_{s} \rightarrow K^{*}_{2}(1430)\overline{\emph{K}}^{*}_{2}(1430))
&=-{G_{F}}(V^{*}_{tb}V_{ts}[({\frac{7}{3}C_{3}+\frac{5}{3}C_{4}-\frac{7}{6}C_{9}-\frac{5}{6}C_{10}})F^{LL,h}_{a}\\
&+2({C_{5}+\frac{1}{3}C_{6}-\frac{1}{2}C_{7}-\frac{1}{6}C_{8}})F^{LR,h}_{a}\\
&+(-\frac{1}{2}C_{8}-\frac{1}{6}C_{7}+C_{6}+\frac{1}{3}C_{5})F^{SP,h}_{a}\\
&+(C_{3}-\frac{1}{2}C_{9})M^{LL,h}_{e}\\
&+(C_{5}-\frac{1}{2}C_{7})M^{LR,h}_{e}\\
&+(C_{3}+2C_{4}-\frac{1}{2}C_{9}-C_{10})M^{LL,h}_{a}\\
&+({2}C_{6}-C_{8})M^{SP,h}_{a}\\
&+(C_{5}-\frac{1}{2}C_{7})M^{LR,h}_{a}])
\end{split}
\end{equation}

\begin{equation}
\begin{split}
A_{h}(B^{0}_{s} \rightarrow f^{'}_{2}(1525)f^{'}_{2}(1525))
&=-{G_{F}}V^{*}_{tb}V_{ts}[({\frac{4}{3}C_{3}+\frac{4}{3}C_{4}-\frac{2}{3}C_{9}-\frac{2}{3}C_{10}})F^{LL,h}_{a}\\
&+({C_{5}+\frac{1}{3}C_{6}-\frac{1}{2}C_{7}-\frac{1}{6}C_{8}})F^{LR,h}_{a}\\
&+({C_{6}+\frac{1}{3}C_{5}-\frac{1}{2}C_{8}-\frac{1}{6}C_{7}})F^{SP,h}_{a}\\
&+(C_{3}+C_{4}-\frac{1}{2}C_{9}-\frac{1}{2}C_{10})M^{LL,h}_{e}\\
&+(C_{5}-\frac{1}{2}C_{7})M^{LR,h}_{e}\\
&+(C_{3}+C_{4}-\frac{1}{2}C_{9}-\frac{1}{2}C_{10})M^{LL,h}_{a}\\
&+(C_{6}-\frac{1}{2}C_{8})M^{SP,h}_{a}\\
&+(C_{5}-\frac{1}{2}C_{7})M^{LR,h}_{a}\\
&+(C_{6}-\frac{1}{2}C_{8})M^{SP,h}_{e}]
\end{split}
\end{equation}

\begin{equation}
\begin{split}
A_{h}(B^{0}_{s} \rightarrow f_{2}(1270)f^{'}_{2}(1525))
&=\frac{G_{F}}{{2}}({V^{*}_{ub}V_{us}[C_{2}M^{LL,h}_{e}]}\\
&-V^{*}_{tb}V_{ts}[({2}C_{4}+\frac{1}{2}C_{10})M^{LL,h}_{e}\\
&+(\frac{1}{2}C_{8}+{2}C_{6})M^{SP,h}_{e}])
\end{split}
\end{equation}

\begin{equation}
\begin{split}
A_{h}(B^{0}_{s} \rightarrow a^{+}_{2}(1320)a^{-}_{2}(1320))
&=\frac{G_{F}}{\sqrt{2}}({V^{*}_{ub}V_{us}[(C_{1}+\frac{1}{3}C_{2})F^{LL,h}_{a}+C_{2}M^{LL,h}_{a}]}\\
&-2V^{*}_{tb}V_{ts}[({{2}C_{3}+\frac{2}{3}C_{4}+{2}C_{5}+\frac{2}{3}C_{6}}\\
&+{\frac{1}{2}C_{7}+\frac{1}{6}C_{8}+\frac{1}{2}C_{9}+\frac{1}{6}C_{10}})F^{LL,h}_{a}\\
&+({2}C_{4}+\frac{1}{2}C_{10})M^{LL,h}_{a}+({2}C_{6}+\frac{1}{2}C_{8})M^{SP,h}_{a}])
\end{split}
\end{equation}

\begin{equation}
\begin{split}
A_{h}(B^{0}_{s} \rightarrow K^{*+}_{2}(1430)\overline{\emph{K}}^{*-}_{2}(1430))
&=\frac{G_{F}}{\sqrt{2}}({V^{*}_{ub}V_{us}[(C_{1}+\frac{1}{3}C_{2})F^{LL,h}_{a}+C_{2}M^{LL,h}_{a}+C_{1}M^{LL,h}_{e}]}\\
&-V^{*}_{tb}V_{ts}[({\frac{7}{3}C_{3}+\frac{5}{3}C_{4}+\frac{1}{3}C_{9}-\frac{1}{3}C_{10}})F^{LL,h}_{a}\\
&+(2{C_{5}+\frac{2}{3}C_{6}+\frac{1}{2}C_{7}+\frac{1}{6}C_{8}})F^{LR,h}_{a}\\
&+(-\frac{1}{2}C_{8}-\frac{1}{6}C_{7}+C_{6}+\frac{1}{3}C_{5})F^{SP,h}_{a}\\
&+(C_{3}+C_{9})M^{LL,h}_{e}\\
&+(C_{5}+C_{7})M^{LR,h}_{e}\\
&+(C_{3}+2C_{4}+\frac{1}{2}C_{9}-C_{10})M^{LL,h}_{a}\\
&+({2}C_{6}+\frac{1}{2}C_{8})M^{SP,h}_{a}\\
&+(C_{5}-\frac{1}{2}C_{7})M^{LR,h}_{a}])
\end{split}
\end{equation}

Based on the  mixing scheme, the helicity amplitudes of $ B^{0}_{s} \rightarrow f_{n}f_{n} (f_{s}f_{s})$ and $ B^{0}_{s} \rightarrow f_{n}f_{s} $ decays are given by~\cite{Jiang:2020eml}

\begin{equation}
\begin{split}
A_{h}(B^{0}_{s} \rightarrow f_{2}(1270)f^{'}_{2}(1525))
&=\sin(2\theta)[A_{h}(B^{0}_{s} \rightarrow f_{n}f_{n})-A_{h}(B^{0}_{s} \rightarrow f_{s}f_{s})]\\
&+\cos(2\theta)A_{h}(B^{0}_{s} \rightarrow f_{n}f_{s}),
\end{split}
\end{equation}
\begin{equation}
\begin{split}
\sqrt{2}A_{h}(B^{0}_{s} \rightarrow (f_{2}(1270)f_{2}(1270))
&=2\cos^2\theta A_{h}(B^{0}_{s} \rightarrow f_{n}f_{n})-\sin(2\theta)A_{h}(B^{0}_{s} \rightarrow f_{n}f_{s})\\
&+2\sin^2\theta A_{h}(B^{0}_{s} \rightarrow f_{s}f_{s}),
\end{split}
\end{equation}
\begin{equation}
\begin{split}
\sqrt{2}A_{h}(B^{0}_{s} \rightarrow (f^{'}_{2}(1525)f^{'}_{2}(1525))
&=2\sin^2\theta A_{h}(B^{0}_{s} \rightarrow f_{n}f_{n})+\sin(2\theta)A_{h}(B^{0}_{s} \rightarrow f_{n}f_{s})\\
&+2\cos^2\theta A_{h}(B^{0}_{s} \rightarrow f_{s}f_{s}).
\end{split}
\end{equation}

In this part, we summarize the functions that appear in the previous sections. For the factorizable annihilation diagrams that the first two diagrams in the Fig.~\ref{fig1}, whose hard scales $t_{i}$ can be written by
\begin{equation}
\begin{split}
&t_{e}=\mathrm{Max}\{{\alpha_{1} M_{B_{s}^{0}}, \beta M_{B_{s}^{0}}, \frac{1}{b_{2}},\frac{1}{b_{3}}\}},\\
&t_{f}=\mathrm{Max}\{{\alpha_{2} M_{B_{s}^{0}}, \beta M_{B_{s}^{0}}, \frac{1}{b_{2}},\frac{1}{b_{3}}\}},\\
&h_{af}(\alpha_{i},\beta, b_{2},  b_{3})=(\frac{\emph{i}{\pi}}{2})^2 H^{(1)}_{0}(\beta M_{B_{s}^{0}}b_{2} )\times[\theta(b_{2}-b_{3})H^{(1)}_{0}(\alpha_{i} M_{B_{s}^{0}} b_{2}) J_{0}(\alpha_{i} M_{B_{s}^{0}} b_{3})\\
&+\theta(b_{3}-b_{2})H^{(1)}_{0}(\alpha_{i} M_{B_{s}^{0}} b_{3}) J_{0}(\alpha_{i} M_{B_{s}^{0}} b_{2})]
S_{t}(x_{3}),
\end{split}
\end{equation}
with\\
\begin{equation}
\begin{split}
&\alpha_{1}=\sqrt {1-(1-r_{2}^2)x_{3}},\\
&\alpha_{2}=\sqrt{[(1-r_{3}^2)x_{2}+r_{3}^2](1-r_{2}^2)},\\
&\beta=\sqrt{[(1-r_{2}^2)(1-x_{3})][r_{3}^2+x_{2}(1-r_{3}^2)]}.
\end{split}
\end{equation}

The parametrized expression of the threshold resummation function $S_{t}(x)$ is~\cite{Li:2001ay}
\renewcommand{\theequation}{{A}.2}
\begin{equation}
\begin{split}
S_{t}(x)=[x(1-x)]^{c}\frac{2^{1+2c}\Gamma(\frac{3}{2}+c)}{\sqrt{\pi}\Gamma(1+c)},
\end{split}
\end{equation}
where $c=0.04$.
The evolution factors $E_{af}(t)$ in the matrix elements are given by
\begin{equation}
\begin{split}
&E_{af}(t)=\alpha_{s}(t)\exp[-S_{2}(t)-S_{3}(t)].
\end{split}
\end{equation}

The Sudakov exponents can be written as
\begin{equation}
\begin{split}
&S_{B}(t)=s(x_{1} \frac{M_{B_{s}^{0}}}{\sqrt{2}},b_{1})+\frac{5}{3}\int^{t}_{1/b_{1}}\emph{d}{\bar{\mu}}\frac{\gamma_{q}(\alpha_{s}({\bar{\mu}}))}{\bar{\mu}},\\
&S_{i}(t)=s(x_{i} \frac{M_{B_{s}^{0}}}{\sqrt{2}},b_{i})+s((1-x_{i}) \frac{M_{B_{s}^{0}}}{\sqrt{2}},b_{i})+2\int^{t}_{1/b}\emph{d}{\bar{\mu}}\frac{\gamma_{q}(\alpha_{s}({\bar{\mu}}))}{\bar{\mu}},
\end{split}
\end{equation}
where the Sudakov factors $s(Q,b)$ are derived from the double logarithmic summation. Its specific expression can be found in the Ref.~\cite{Keum:2000wi}.
For the nonfactorizable contribution, due to the small numerical effect, we drop  the threshold resummation function~\cite{Li:2002mi}.
For the nonfactorizable annihilation diagrams, the scales and the hard functions are~\cite{Beuzit:2001pt}
\begin{equation}
\begin{split}
&t_{g}=\mathrm{Max}\{{\alpha  M_{B_{s}^{0}}, \sqrt{|\beta_{1}|} M_{B_{s}^{0}}, \frac{1}{b_{1}},\frac{1}{b_{2}}\}},\\
&t_{h}=\mathrm{Max}\{{\alpha  M_{B_{s}^{0}}, \sqrt{|\beta_{2}|} M_{B_{s}^{0}}, \frac{1}{b_{1}},\frac{1}{b_{2}}\}},\\
&E_{anf}(t)=\alpha_{s}(t)\exp[-S_{B}(t)-S_{2}(t)-S_{3}(t)]|_{b_{2}=b_{3}},\\
&h_{anf}(\alpha,\beta_{i}, b_{1}, b_{2})=\frac{\emph{i}{\pi}}{2}[\theta(b_{1}-b_{2})H^{(1)}_{0}(\alpha M_{B_{s}^{0}} b_{1}) J_{0}(\alpha M_{B_{s}^{0}} b_{2})\\
&+\theta(b_{2}-b_{1})H^{(1)}_{0}(\alpha M_{B_{s}^{0}} b_{2}) J_{0}(\alpha M_{B_{s}^{0}} b_{1})]\\
&\quad \quad \quad \quad\quad\quad\quad\quad\quad\times \begin{cases}
\frac{\emph{i}{\pi}}{2} H^{(1)}_{0}(M_{B_{s}^{0}}b_{1}\sqrt{|\beta_{i}|}), & {\beta_{i} < 0},\\
K_{0}(M_{B_{s}^{0}}b_{1}\sqrt{\beta_{i}}), & {\beta_{i} > 0},\\
\end{cases}\\
\end{split}
\end{equation}
with $i=1,2$\\
\begin{equation}
\begin{split}
&\alpha=\sqrt {(1-x_{3})(1-r_{2}^2)[r_{3}^2+x_{2}(1-r_{3}^2)]},\\
&\beta_{1}=1-[(1-r_{3}^2)(1-x_{2})-x_{1}][r_{2}^2+x_{3}(1-r_{2}^2)],\\
&\beta_{2}=(1-r_{2}^2)(1-x_{3})[x_{1}-x_{2}(1-r_{3}^2)-r_{3}^2].
\end{split}
\end{equation}

For the nonfactorizable emission diagrams, the rest functions are follows
\begin{equation}
\begin{split}
&t_{c}=\mathrm{Max}\{M_{B_{s}^{0}}\sqrt{(1-r_{2}^2)x_{1}x_{3}}, M_{B_{s}^{0}}\sqrt{|[(1-r_{3}^2)(x_{2}-1)+x_{1}][r_{2}^2+x_{3}(1-r_{2}^2)]|},\frac{1}{b_{1}}, \frac{1}{b_{2}}\},\\
&t_{d}=\mathrm{Max}\{M_{B_{s}^{0}}\sqrt{(1-r_{2}^2)x_{1}x_{3}}, M_{B_{s}^{0}}\sqrt{|[(r_{3}^2-1)x_{2}+x_{1}]x_{3}(1-r_{2}^2)|},\frac{1}{b_{1}}, \frac{1}{b_{2}}\}.
\end{split}
\end{equation}

The function $h_{enf}$ can be determined by
\begin{equation}
\begin{split}
&h_{enf}(\alpha,\beta_{i}, b_{1},  b_{2})=  [\theta(b_{2}-b_{1})I_{0}(\alpha b_{1})K_{0}(\alpha b_{2}) +\theta(b_{1}-b_{2})I_{0}(\alpha b_{2})K_{0}(\alpha b_{1})]\\
&\quad \quad \quad \quad\quad\quad\quad\quad\quad\times \begin{cases}
\frac{\emph{i}{\pi}}{2} H^{(1)}_{0}(M_{B_{s}^{0}}b_{2}\sqrt{|\beta^{2}_{i}|}), & {\beta^{2}_{i} < 0},\\
K_{0}(M_{B_{s}^{0}}b_{2}\beta_{i}), & {\beta^{2}_{i} > 0},\\
\end{cases}\\
\end{split}
\end{equation}
with $i=1,2$\\
\begin{equation}
\begin{split}
&E_{enf}(t)=\alpha_{s}(t)\exp[-S_{B}(t)-S_{2}(t)-S_{3}(t)]|_{b_{1}=b_{3}},\\
&\alpha=M_{B_{s}^{0}}\sqrt {(1-r_{2}^2)x_{1}x_{3}},\\
&\beta^{2}_{1}=[(1-r_{3}^2)(x_{2}-1)+x_{1}][r_{2}^2+x_{3}(1-r_{2}^2)],\\
&\beta^{2}_{2}=[(r_{3}^2-1)x_{2}+x_{1}]x_{3}(1-r_{2}^2).
\end{split}
\end{equation}


\end{document}